\documentclass[11pt]{article}
\usepackage[totalwidth=460truept,totalheight=600truept]{geometry}
\usepackage{latexsym,amssymb,amsmath,graphicx}
\usepackage[T1]{fontenc}

\def\theequation{\arabic{section}.\arabic{equation}}

\renewcommand{\theequation}{\thesection.\arabic{equation}}
\global\arraycolsep=1truept

\begin{document}

\hfill IFUP-TH 2008/17

\vskip 1.4truecm

\begin{center}
{\huge \textbf{Weighted Power Counting}}

{\huge \textbf{\large \vskip .1truecm}}

{\huge \textbf{And Lorentz Violating Gauge Theories. }}

{\huge \textbf{\large \vskip .1truecm}}

{\huge \textbf{I: General Properties}}

\vskip 1.5truecm

\textsl{Damiano Anselmi}

\textit{Dipartimento di Fisica ``Enrico Fermi'', Universit\`{a} di Pisa, }

\textit{Largo Pontecorvo 3, I-56127 Pisa, Italy, }

\textit{and INFN, Sezione di Pisa, Pisa, Italy}

damiano.anselmi@df.unipi.it

\vskip 2truecm

\textbf{Abstract}
\end{center}

\bigskip

{\small We construct local, unitary gauge theories that violate Lorentz
symmetry explicitly at high energies and are renormalizable by weighted
power counting. They contain higher space derivatives, which improve the
behavior of propagators at large momenta, but no higher time derivatives. We
show that the regularity of the gauge-field propagator privileges a
particular spacetime breaking, the one into into space and time. We then
concentrate on the simplest class of models, study four dimensional examples
and discuss a number of issues that arise in our approach, such as the
low-energy recovery of Lorentz invariance.}

\vskip 1truecm

\vfill\eject

\section{Introduction}

\setcounter{equation}{0}

Lorentz symmetry has been verified in many experiments with great precision 
\cite{koste}. However, different types of arguments have lead some authors
to suggest that it could be violated at very high energies \cite
{colective,kostelecky,colective2}. This possibility has raised a
considerable interest, because, if true, it would substantially affect our
understanding of Nature. The Lorentz violating extension of the Standard
Model \cite{kostelecky} contains a large amount of new parameters. Bounds on
many of them, particularly those belonging to the power-counting
renormalizable subsector, are available. Their updated values are reported
in ref. \cite{koste2}.

In quantum field theory, the classification of local, unitarity, polynomial
and renormalizable models changes dramatically if we do not assume that
Lorentz invariance is exact at arbitrarily high energies \cite
{renolor,confnolor}. In that case, higher space derivatives are allowed and
can improve the behavior of propagators at large momenta. A number of
theories that are not renormalizable by ordinary power counting become
renormalizable in the framework of a ``weighted power counting'' \cite
{renolor}, which assigns different weights to space and time, and ensures
that no term containing higher time derivatives is generated by
renormalization, in agreement with unitarity. Having studied scalar and
fermion theories in ref.s \cite{renolor,confnolor}, here we begin the study
of gauge theories, focusing on the simplest class of models. The
investigation is completed in a second paper \cite{LVgauge1suAbar}, to which
we refer as paper II, which contains the classification of renormalizable
gauge theories.

The theories we are interested in must be local and polynomial, free of
infrared divergences in Feynman diagrams at non-exceptional external
momenta, and renormalizable by weighted power counting. We find that in the
presence of gauge interactions the set of renormalizable theories is more
restricted than in the scalar-fermion framework. Due to the particular
structure of the gauge-field propagator, Feynman diagrams are plagued with
certain spurious subdivergences. We are able to prove that they cancel out
when spacetime is broken into space and time, and certain other restrictions
are fulfilled.

A more delicate physical issue is the low-energy recovery of Lorentz
symmetry. Once Lorentz symmetry is violated at high energies, its low-energy
recovery is not guaranteed, because renormalization makes the low-energy
parameters run independently. One possibility is that the Lorentz invariant
surface is RG stable \cite{nielsen}, otherwise a suitable fine-tuning must
be advocated.

In other domains of physics, such as the theory of critical phenomena, where
Lorentz symmetry is not a fundamental requirement, certain scalar models of
the types classified in ref. \cite{renolor} have already been studied \cite
{lifshitz} and have physical applications.

The paper is organized as follows. In section 2 we review the weighted power
counting for scalar-fermion theories. In section 3 we extend it to Lorentz
violating gauge theories and define the class of models we focus on in this
paper. We study the conditions for renormalizability, absence of infrared
divergences in Feynman diagrams and regularity of the propagator. In section
4 we prove that the theories are renormalizable to all orders, using the
Batalin-Vilkovisky formalism. In section 5 we study four dimensional
examples and the low-energy recovery of Lorentz invariance. In section 6 we
discuss strictly renormalizable and weighted scale invariant theories. In
section 7 we study the Proca Lorentz violating theories, and prove that they
are not renormalizable. Section 8 contains our conclusions. In appendix A we
classify the quadratic terms of the gauge-field lagrangian and in appendix B
we derive sufficient conditions for the absence of spurious subdivergences.

\section{Weighted power counting}

\setcounter{equation}{0}

In this section we briefly review the weighted power counting criterion of
refs. \cite{renolor,confnolor}. The simplest framework to study Lorentz
violations is to assume that the $d$-dimensional spacetime manifold $M=%
\mathbb{R}^{d}$ is split into the product $\hat{M}\times \bar{M}$ of two
submanifolds, a $\hat{d}$-dimensional submanifold $\hat{M}=\mathbb{R}^{\hat{d%
}}$, containing time and possibly some space coordinates, and a $\bar{d}$%
-dimensional space submanifold $\bar{M}=\mathbb{R}^{\bar{d}}$, so that the $%
d $-dimensional Lorentz group $O(1,d-1)$ is broken to a residual Lorentz
group $O(1,\hat{d}-1)\times O(\bar{d})$. In this paper we study
renormalization in this simplified framework. The generalization to the most
general breaking is done in paper II.

The partial derivative $\partial $ is decomposed as $(\hat{\partial},\bar{%
\partial})$, where $\hat{\partial}$ and $\bar{\partial}$ act on the
subspaces $\hat{M}$ and $\bar{M}$, respectively. Coordinates, momenta and
spacetime indices are decomposed similarly. Consider a free scalar theory
with (Euclidean) lagrangian 
\begin{equation}
\mathcal{L}_{\hbox{free}}=\frac{1}{2}(\hat{\partial}\varphi )^{2}+\frac{1}{%
2\Lambda _{L}^{2n-2}}(\bar{\partial}^{n}\varphi )^{2},  \label{free}
\end{equation}
where $\Lambda _{L}$ is an energy scale and $n$ is an integer $>1$. Up to
total derivatives it is not necessary to specify how the $\bar{\partial}$'s
are contracted among themselves. The coefficient of $(\bar{\partial}%
^{n}\varphi )^{2}$ must be positive to have a positive energy in the
Minkowskian framework. The theory (\ref{free}) is invariant under the
weighted rescaling 
\begin{equation}
\hat{x}\rightarrow \hat{x}\ \mathrm{e}^{-\Omega },\qquad \bar{x}\rightarrow 
\bar{x}\ \mathrm{e}^{-\Omega /n},\qquad \varphi \rightarrow \varphi \ 
\mathrm{e}^{\Omega (\text{\dj }/2-1)},  \label{scale}
\end{equation}
where \dj $=\hat{d}+\bar{d}/n$ is the ``weighted dimension''. Note that $%
\Lambda _{L}$ is not rescaled.

The interacting theory is defined as a perturbative expansion around the
free theory (\ref{free}). For the purposes of renormalization, the masses
and the other quadratic terms can be treated perturbatively, since the
counterterms depend polynomially on them. Denote the ``weight'' of an object 
$\mathcal{O}$ by $[\mathcal{O}]$ and assign weights to coordinates, momenta
and fields as follows: 
\begin{equation}
\lbrack \hat{x}]=-1,\qquad [\bar{x}]=-\frac{1}{n},\qquad [\hat{\partial}%
]=1,\qquad [\bar{\partial}]=\frac{1}{n},\qquad [\varphi ]=\frac{\text{\dj }}{%
2}-1,  \label{weights}
\end{equation}
while $\Lambda _{L}$ is weightless. Polynomiality demands that the weight of 
$\varphi $ be strictly positive, so we assume \dj $>2$.

We say that $P_{k,n}(\hat{p},\bar{p})$ is a weighted polynomial in $\hat{p}$
and $\bar{p}$, of degree $k$, where $k$ is a multiple of $1/n$, if $%
P_{k,n}(\xi ^{n}\hat{p},\xi \bar{p})$ is a polynomial of degree $kn$ in $\xi 
$. A diagram $G$ with $L$ loops, $V$ vertices and $I$ internal legs gives an
integral of the form 
\[
\mathcal{I}_{G}(k)=\int \frac{\mathrm{d}^{dL}p}{(2\pi )^{d}}\prod_{i=1}^{I}%
\mathcal{P}_{i}(p,k)\prod_{j=1}^{V}\mathcal{V}_{j}(p,k), 
\]
where $p$ are the loop momenta, $k$ are the external momenta, $\mathcal{P}%
_{i}(p,k)$ are the propagators and $\mathcal{V}_{j}(p,k)$ are the vertices.
The momentum integration measure $\mathrm{d}^{d}p$ has weight \dj . The
propagator is equal to $1$ divided by a weighted polynomial of degree $2$.
We can assume that, as far as their momentum dependence is concerned, the
vertices are weighted monomials of certain degrees $\delta _{j}$. Rescaling $%
k$ and as $(\hat{k},\bar{k})\rightarrow (\lambda \hat{k},\lambda ^{1/n}\bar{k%
})$, the integral $\mathcal{I}_{G}(k)$ rescales with a factor equal to its
total weight $\omega (G)$. By locality, the divergent part of $\mathcal{I}%
_{G}(k)$ is a weighted polynomial of degree $\omega (G)$. Assume that the
lagrangian contains all vertices that have weights not greater than \dj\ and
only those. This bound excludes terms with higher time derivatives. Then we
find 
\begin{equation}
\omega (G)\leq \text{\dj }-E_{s}\frac{\text{\dj }-2}{2},  \label{abbound}
\end{equation}
where $E_{s}$ is the number of external scalar legs. Formula (\ref{abbound})
and \dj $>2$ ensure that every counterterm has a weight not larger than \dj
, therefore it can be subtracted renormalizing the fields and couplings of
the lagrangian, and no new vertex needs to be introduced.

The lagrangian terms of weight \dj\ are strictly renormalizable, those of
weights smaller than \dj\ super-renormalizable and those of weights greater
than \dj\ non-renormalizable. The weighted power counting criterion amounts
to demand that the theory contains no parameter of negative weight.

Simple examples of renormalizable theories are the $\varphi ^{4}$, 
\hbox{\dj
}$=4$ models 
\begin{equation}
\mathcal{L}_{\hbox{\dj }=4}=\frac{1}{2}(\widehat{\partial }\varphi )^{2}+%
\frac{1}{2\Lambda _{L}^{2(n-1)}}(\overline{\partial }^{n}\varphi )^{2}+\frac{%
\lambda }{4!\Lambda _{L}^{d-4}}\varphi ^{4}  \label{scal}
\end{equation}
and the $\varphi ^{6}$, \hbox{\dj }$=3$ models 
\begin{equation}
\mathcal{L}_{\hbox{\dj }=3}=\frac{1}{2}(\widehat{\partial }\varphi )^{2}+%
\frac{1}{2\Lambda _{L}^{2(n-1)}}(\overline{\partial }^{n}\varphi )^{2}+\frac{%
1}{4!\Lambda _{L}^{2(n-1)}}\sum_{\alpha }\lambda _{\alpha }\left[ {\overline{%
\partial }}^{n}{\varphi }^{4}\right] _{\alpha }+\frac{\lambda _{6}}{%
6!\Lambda _{L}^{2(n-1)}}\varphi ^{6}.  \label{even}
\end{equation}
where $\left[ {\overline{\partial }}^{n}{\varphi }^{4}\right] _{\alpha }$
denotes a basis of inequivalent terms constructed with $n$ derivatives ${%
\overline{\partial }}${\ acting on four }${\varphi }$'s\footnote{%
Because of $O(\overline{d})$-invariance, these exist no such terms if $n$ is
odd.}. Only the strictly-renormalizable terms have been listed in (\ref{scal}%
) and (\ref{even}). It is straightforward to complete the actions adding the
super-renormalizable terms, which are those that contain fewer derivatives
and/or $\varphi $-powers.

The considerations just recalled are easily generalized to fermions. The
weight of a fermion field is (\dj $-1)/2$, so polynomiality is ensured,
because \dj\ is necessarily greater than 1 (if $d>1$). Formula (\ref{abbound}%
) becomes 
\[
\omega (G)\leq \text{\dj }-E_{s}\frac{\text{\dj }-2}{2}-E_{f}\frac{\text{\dj 
}-1}{2}, 
\]
where $E_{f}$ is the number of external fermionic legs.

Our investigation focuses on the theories that do contain higher space
derivatives ($n\geq 2$). Indeed, the theories with $n=1$, which can be
either Lorentz invariant or Lorentz violating, obey the usual rules of power
counting.

\section{Lorentz violating gauge theories}

\setcounter{equation}{0}

Having decomposed the partial derivative operator as $\partial =(\hat{%
\partial},\bar{\partial})$, the gauge field has to be decomposed similarly.
We write $A^{\prime }=(\hat{A}^{\prime },\bar{A}^{\prime })\equiv gA=g(\hat{A%
},\bar{A})$, where $g$ is the gauge coupling and $A_{\mu }=A_{\mu }^{a}T^{a}$%
, with $T^{a}$ anti-Hermitian. The covariant derivative is decomposed as 
\begin{equation}
D=(\hat{D},\bar{D})=(\hat{\partial}+\hat{A}^{\prime },\bar{\partial}+\bar{A}%
^{\prime }).  \label{3a}
\end{equation}
With the weight assignments 
\[
\lbrack \hat{A}^{\prime }]=[\hat{D}]=1,\qquad [\bar{A}^{\prime }]=[\bar{D}]=%
\frac{1}{n}, 
\]
the decomposition (\ref{3a}) is compatible with the weighted rescaling. The
field strength is split into three sets of components, namely 
\begin{equation}
\hat{F}_{\mu \nu }\equiv F_{\hat{\mu}\hat{\nu}},\qquad \tilde{F}_{\mu \nu
}\equiv F_{\hat{\mu}\bar{\nu}},\qquad \bar{F}_{\mu \nu }\equiv F_{\bar{\mu}%
\bar{\nu}}.  \label{3b}
\end{equation}
Since the kinetic lagrangian must contain $(\hat{\partial}\hat{A})^{2}$, the
weight of $\hat{A}$ is \dj $/2-1$, hence $[g]=2-$\dj $/2$. We can read $[%
\bar{A}]$ from $[\tilde{F}]=[\bar{\partial}]+[\hat{A}]=[\hat{\partial}]+[%
\bar{A}]$. In summary, 
\begin{equation}
\lbrack \hat{A}]=\frac{\text{\dj }}{2}-1,\qquad [\bar{A}]=\frac{\text{\dj }}{%
2}-2+\frac{1}{n},\qquad [\hat{F}]=\frac{\text{\dj }}{2},\qquad [\tilde{F}]=%
\frac{\text{\dj }}{2}-1+\frac{1}{n},\qquad [\bar{F}]=\frac{\text{\dj }}{2}-2+%
\frac{2}{n}.  \label{we}
\end{equation}
Since the weight of $g$ cannot be negative, we must have 
\begin{equation}
\text{\dj }\leq 4.  \label{ono}
\end{equation}

In this paper we focus on the ``$1/\alpha $ theories'', namely those that
have a lagrangian of the form 
\begin{equation}
\mathcal{L}=\frac{1}{\alpha }\mathcal{L}_{r}(gA,g\varphi ,g\psi ,g\bar{C}%
,gC,\lambda ).  \label{sur}
\end{equation}
Here $C$ and $\bar{C}$ are the ghosts and antighosts, $\varphi $ are the
scalar fields and $\psi $ are the fermions. Moreover, the reduced lagrangian 
$\mathcal{L}_{r}$ depends polynomially on $g$ and the other parameters $%
\lambda $, and $[\lambda ]\geq 0$. The renormalizability of the structure (%
\ref{sur})\ is easy to prove (see (\ref{ax})).

When \dj $=4$ the gauge coupling $g$ is weightless and the theory can always
be written in the $1/\alpha $ form with a suitable redefinition of
parameters. Instead, when \dj $<4$ the $1/\alpha $ theories are a small
subset of the allowed theories. For example, in $d=4$ the weights of $%
\mathcal{L}_{r}$, $\hat{\partial}$, $g\hat{A}$, $g\varphi $, $g\bar{C}$, $gC$
and $g\psi $ coincide with their dimensions in units of mass, and only the
weights of $\bar{\partial}$ and $g\bar{A}$, which are equal to $1/n$, differ
from their dimensions. Then the lagrangian contains just the usual
power-counting renormalizable terms, plus the terms that can be constructed
with $\bar{D}$, $\bar{F}$ and $\tilde{F}$. The form of the lagrangian does
not change in $d\neq 4$. Polynomiality is always ensured.

Even if the $1/\alpha $ theories are not particularly interesting from the
physical point of view, it is convenient to start from them, because the
simplified structure (\ref{sur}) allows us to illustrate the basic
properties of Lorentz violating gauge theories without unnecessary
complicacies. The most general case is studied in paper II.

Observe that the theories (\ref{sur}) cannot contain higher time
derivatives, as desired. Indeed, by $O(1,\hat{d}-1)$-invariance a term with
three $\hat{\partial}$'s in $\mathcal{L}_{r}$ must contain at least another $%
\hat{\partial}$, or a $g\hat{A}$, or a fermion bilinear such as $g^{2}\bar{%
\psi}\hat{\gamma}\psi $. However, the weights of $\hat{\partial}^{4}$ and $g%
\hat{\partial}^{3}\hat{A}$ are already equal to four, so no other leg can be
attached to such objects, and the weight of $g^{2}\hat{\partial}^{3}\bar{\psi%
}\hat{\gamma}\psi $ is equal to six.

It is convenient to write the action 
\begin{equation}
\mathcal{S}_{0}=\int \mathrm{d}^{d}x\left( \mathcal{L}_{Q}+\mathcal{L}%
_{I}\right) \equiv \mathcal{S}_{Q}+\mathcal{S}_{I},  \label{s0}
\end{equation}
as the sum of two gauge-invariant contributions, the quadratic terms $%
\mathcal{S}_{Q}$ plus the vertex terms $\mathcal{S}_{I}$. By ``quadratic
terms'' we mean the terms constructed with two field strengths and possibly
covariant derivatives. By ``vertex terms'' we mean the terms constructed
with at least three field strengths, and possibly covariant derivatives.

In Appendix A we prove that, up to total derivatives, the quadratic part $%
\mathcal{L}_{Q}$ of the lagrangian reads (in the Euclidean framework) 
\begin{equation}
\mathcal{L}_{Q}=\frac{1}{4}\left\{ F_{\hat{\mu}\hat{\nu}}^{2}+2F_{\hat{\mu}%
\bar{\nu}}\eta (\bar{\Upsilon})F_{\hat{\mu}\bar{\nu}}+F_{\bar{\mu}\bar{\nu}%
}\tau (\bar{\Upsilon})F_{\bar{\mu}\bar{\nu}}+\frac{1}{\Lambda _{L}^{2}}(D_{%
\hat{\rho}}F_{\bar{\mu}\bar{\nu}})\xi (\bar{\Upsilon})(D_{\hat{\rho}}F_{\bar{%
\mu}\bar{\nu}})\right\} .  \label{l0}
\end{equation}
Here $\bar{\Upsilon}\equiv -\bar{D}^{2}/\Lambda _{L}^{2}$ and $\eta $, $\tau 
$ and $\xi $ are polynomials of degrees $n-1$, $2n-2$ and $n-2$,
respectively. We have expansions 
\begin{equation}
\eta (\bar{\Upsilon})=\sum_{i=0}^{n-1}\eta _{n-1-i}\bar{\Upsilon}^{i},\qquad
[\eta _{j}]=\frac{2j}{n},  \label{expo}
\end{equation}
and similar, where $\eta _{i}$ are dimensionless constants of non-negative
weights.

In momentum space we see that the free action is positive definite if and
only if 
\begin{equation}
\eta >0,\qquad \tilde{\eta}\equiv \eta +\frac{\bar{k}^{2}}{\Lambda _{L}^{2}}%
\xi >0,\qquad \tau >0,  \label{pos}
\end{equation}
where now $\eta $, $\tau $ and $\xi $ are functions of $\bar{k}^{2}/\Lambda
_{L}^{2}$.

In the parametrization (\ref{l0}) the scale $\Lambda _{L}$ is a redundant
parameter. It is mainly used to match the dimensions in units of mass, so
that the other parameters (e.g. the $\eta _{j}$'s) can be assumed to be
dimensionless, but possibly weightful. The $\Lambda _{L}$-redundancy implies
that $\Lambda _{L}$ is RG invariant, so its beta function vanishes by
definition.

\paragraph{BRST\ symmetry and gauge fixing}

The usual BRST\ symmetry \cite{brs} 
\begin{eqnarray*}
sA_{\mu }^{a} &=&D_{\mu }^{ab}C^{b}=\partial _{\mu }C^{a}+gf^{abc}A_{\mu
}^{b}C^{c},\qquad sC^{a}=-\frac{g}{2}f^{abc}C^{b}C^{c}, \\
s\bar{C}^{a} &=&B^{a},\qquad sB^{a}=0,\qquad s\psi
^{i}=-gT_{ij}^{a}C^{a}\psi ^{j},
\end{eqnarray*}
etc., where $B^{a}$ are Lagrange multipliers for the gauge-fixing, is
automatically compatible with the weighted power counting. The quadratic
terms of the ghost Lagrangian contain $\bar{C}\hat{\partial}^{2}C$ and $%
B^{2} $, and have weight \dj , so we have the weight assignments 
\begin{equation}
\lbrack C]=[\bar{C}]=\frac{\text{\dj }}{2}-1,\qquad [s]=1,\qquad [B]=\frac{%
\text{\dj }}{2}.  \label{g1}
\end{equation}

The most convenient gauge-fixing $\mathcal{G}^{a}$ is linear in the gauge
potential and gives 
\begin{equation}
\mathcal{L}_{\text{gf}}=s\Psi ,\qquad \Psi =\bar{C}^{a}\left( -\frac{\lambda 
}{2}B^{a}+\mathcal{G}^{a}\right) ,\qquad \mathcal{G}^{a}\equiv \hat{\partial}%
\cdot \hat{A}^{a}+\zeta \left( \bar{\upsilon}\right) \bar{\partial}\cdot 
\bar{A}^{a},  \label{gf}
\end{equation}
where $\lambda $ is a dimensionless, weightless constant, $\bar{\upsilon}%
\equiv -\bar{\partial}^{2}/\Lambda _{L}^{2}$ and $\zeta $ is a polynomial of
degree $n-1$. We demand 
\begin{equation}
\zeta >0,  \label{gfpos}
\end{equation}
to include the ``Coulomb gauge-fixing'' $\bar{\partial}\cdot \bar{A}^{a}$.

The total gauge-fixed action is finally 
\begin{equation}
\mathcal{S}=\int \mathrm{d}^{d}x\left( \mathcal{L}_{Q}+\mathcal{L}_{I}+%
\mathcal{L}_{\text{gf}}\right) \equiv \mathcal{S}_{0}+\mathcal{S}_{\text{gf}%
}.  \label{basis}
\end{equation}

\paragraph{Propagator}

The (Euclidean) gauge-field propagator can be worked out from the free
subsector of (\ref{basis}), after integrating $B^{a}$ out, which amounts to
add 
\begin{equation}
\frac{1}{2\lambda }(\mathcal{G}^{a})^{2}  \label{gu}
\end{equation}
to $\mathcal{L}_{Q}$. The result is\footnote{%
A similar propagator, in a different context, has already appeared in ref. 
\cite{regula3d2}.} 
\begin{equation}
\langle A(k)\ A(-k)\rangle =\left( 
\begin{array}{cc}
\langle \hat{A}\hat{A}\rangle & \langle \hat{A}\bar{A}\rangle \\ 
\langle \bar{A}\hat{A}\rangle & \langle \bar{A}\bar{A}\rangle
\end{array}
\right) =\left( 
\begin{array}{cc}
u\hat{\delta}+s\hat{k}\hat{k} & r\hat{k}\bar{k} \\ 
r\bar{k}\hat{k} & v\bar{\delta}+t\bar{k}\bar{k}
\end{array}
\right) ,  \label{pros}
\end{equation}
with 
\begin{eqnarray*}
u &=&\frac{1}{D(1,\eta )},\qquad s=\frac{\lambda }{D^{2}(1,\zeta )}+\frac{-%
\hat{k}^{2}+\zeta \left( \frac{\zeta }{\eta }-2\right) \bar{k}^{2}}{D(1,\eta
)D^{2}(1,\zeta )},\qquad r=\frac{\lambda -\frac{\zeta }{\eta }}{%
D^{2}(1,\zeta )}, \\
v &=&\frac{1}{D(\tilde{\eta},\tau )},\qquad t=\frac{\lambda }{D^{2}(1,\zeta )%
}+\frac{\left( \frac{\tilde{\tau}}{\eta }-2\zeta \right) \hat{k}^{2}-\zeta
^{2}\bar{k}^{2}}{D(\tilde{\eta},\tau )D^{2}(1,\zeta )},
\end{eqnarray*}
where 
\[
D(x,y)\equiv x\hat{k}^{2}+y\bar{k}^{2},\qquad \tilde{\eta}=\eta +\frac{\bar{k%
}^{2}}{\Lambda _{L}^{2}}\xi ,\qquad \tilde{\tau}=\tau +\frac{\hat{k}^{2}}{%
\Lambda _{L}^{2}}\xi , 
\]
and now $\eta $, $\tau $, $\xi $ and $\zeta $, as well as $x$ and $y$, are
meant as functions of $\bar{k}^{2}/\Lambda _{L}^{2}$. The ghost propagator
is 
\begin{equation}
\frac{1}{D(1,\zeta )}.  \label{prosg}
\end{equation}

A simple gauge choice (``Feynman gauge'') is 
\begin{equation}
\lambda =1,\qquad \zeta =\eta .  \label{fg}
\end{equation}
Then, both $\langle \hat{A}\bar{A}\rangle $ and $s$ vanish, so 
\begin{equation}
u=\frac{1}{D(1,\eta )},\qquad s=r=0,\qquad v=\frac{1}{D(\tilde{\eta},\tau )}%
,\qquad t=\frac{\tilde{\tau}-\eta ^{2}}{\eta D(\tilde{\eta},\tau )D(1,\eta )}%
.  \label{fg2}
\end{equation}

\paragraph{Physical degrees of freedom and dispersion relations}

To study the physical degrees of freedom we choose the Coulomb gauge-fixing 
\[
\mathcal{G}_{\mathrm{C}}^{a}=\bar{\partial}\cdot \bar{A}^{a}. 
\]
It can be reached from the more general gauge-fixing (\ref{gf}) taking the
limit $\lambda \rightarrow \infty $, $\zeta \rightarrow \infty $ in (\ref{gf}%
), (\ref{gu}) and (\ref{pros}), with $\varsigma \equiv \lambda /\zeta ^{2}$
fixed, and rescaling the antighosts and the Lagrange multiplier as $\bar{C}%
^{a}\rightarrow \bar{C}^{a}/\zeta $, $B^{a}\rightarrow B^{a}/\zeta $. The
quadratic lagrangian $\mathcal{L}_{Q}+(\bar{\partial}\cdot \bar{A}%
^{a})^{2}/(2\varsigma )$ gives the propagators 
\begin{eqnarray*}
\langle \hat{A}(k)\ \hat{A}(-k)\rangle &=&\frac{1}{D(1,\eta )}\left( \hat{%
\delta}+\frac{\hat{k}\hat{k}}{\bar{k}^{2}\eta }\right) +\frac{\varsigma \hat{%
k}\hat{k}}{(\bar{k}^{2})^{2}},\qquad \langle \hat{A}(k)\ \bar{A}(-k)\rangle =%
\frac{\varsigma \hat{k}\bar{k}}{(\bar{k}^{2})^{2}}, \\
\langle \bar{A}(k)\ \bar{A}(-k)\rangle &=&\frac{1}{D(\tilde{\eta},\tau )}%
\left( \bar{\delta}-\frac{\bar{k}\bar{k}}{\bar{k}^{2}}\right) +\frac{%
\varsigma \bar{k}\bar{k}}{(\bar{k}^{2})^{2}}.
\end{eqnarray*}
Writing $\hat{k}=(iE,\mathbf{\hat{k}})$ and studying the poles, we see that
the $\bar{A}$-sector propagates $\bar{d}-1$ degrees of freedom with energies 
\[
E=\sqrt{\mathbf{\hat{k}}^{2}+\bar{k}^{2}\frac{\tau (\bar{k}^{2}/\Lambda
_{L}^{2})}{\tilde{\eta}(\bar{k}^{2}/\Lambda _{L}^{2})}}, 
\]
while the $\hat{A}$-sector propagates $\hat{d}-1$ degrees of freedom with
energies 
\[
E=\sqrt{\mathbf{\hat{k}}^{2}+\bar{k}^{2}\eta (\bar{k}^{2}/\Lambda _{L}^{2})}%
. 
\]
Indeed, the matrix $\hat{\delta}+\hat{k}\hat{k}/(\bar{k}^{2}\eta )$ has one
null eigenvector on the pole, since its determinant is equal to $D(1,\eta )/(%
\bar{k}^{2}\eta )$. The residues are positive in the Minkowskian framework.
This can be immediately seen using $SO(1,\hat{d}-1)$ invariance to set $%
\mathbf{\hat{k}}=0$ (at $\bar{k}\neq 0$) and $SO(\bar{d})$ invariance to set
all $\bar{k}$-components but one to zero.

Finally, the ghost propagator becomes $1/\bar{k}^{2}$, which has no pole. In
total, the physical degrees of freedom are $d-2$, as expected.

\paragraph{Regularity of the propagator}

A propagator is regular if it is the ratio 
\begin{equation}
\frac{P_{r}(\hat{k},\bar{k})}{P_{2s}^{\prime }(\hat{k},\bar{k})}
\label{uvbeha}
\end{equation}
of two weighted polynomials of degrees $r$ and $2s$, where $r$ and $s$ are
integers, such that the denominator $P_{2s}^{\prime }(\hat{k},\overline{k})$
is non-negative (in the Euclidean framework), non-vanishing when either $%
\hat{k}\neq 0$ or $\overline{k}\neq 0$, and has the form 
\begin{equation}
P_{s}^{\prime }(\hat{k},\bar{k})=\hat{\omega}(\hat{k}^{2})^{s}+\bar{\omega}(%
\bar{k}^{2})^{ns}+\cdots ,  \label{ba}
\end{equation}
with $\hat{\omega}>0$, $\bar{\omega}>0$, where the dots collect the terms $(%
\hat{k}^{2})^{j-m}(\bar{k}^{2})^{mn}$ with $j<s$, $0\leq m\leq j$, and $j=s$%
, $0<m<s$.

The regularity conditions just stated ensure that: $a$) the derivatives with
respect to $\hat{k}$ improve the large-$\bar{k}$ behavior (because $\bar{%
\omega}\neq 0$), besides the large-$\hat{k}$ and overall ones; and $b$) the
derivatives with respect to $\bar{k}$ improve the large-$\hat{k}$ behavior
(because $\hat{\omega}\neq 0$), besides the large-$\bar{k}$ and overall
ones. The overall divergences of the $\hat{k}$-subintegrals are local in $%
\bar{k}$ and the overall divergences of the $\bar{k}$-subintegrals are local
in $\hat{k}$ (once subdiagrams have been inductively subtracted).

In this paper we use the dimensional-regularization technique. We recall 
\cite{renolor} that it is necessary to continue both $\hat{d}$ and $\bar{d}$
to complex values, say $\hat{d}-\varepsilon _{1}$ and $\bar{d}-\varepsilon
_{2}$, respectively. In the framework of the dimensional regularization the
absence of $\hat{k}$- and $\bar{k}$-subdivergences is immediate to prove:\
being local, the $\hat{k}$-subdivergences are killed by the (dimensionally
continued) $\bar{k}$-subintegrals and the $\bar{k}$-subdivergences are
killed by the $\hat{k}$-subintegrals. More explicitly, at one loop we have
integrals of the form 
\[
\int \frac{\mathrm{d}^{\hat{d}-\varepsilon _{1}}\hat{k}}{(2\pi )^{\hat{d}}}%
\left[ \int \frac{\mathrm{d}^{\bar{d}-\varepsilon _{2}}\bar{k}}{(2\pi )^{%
\bar{d}}}\frac{V(\hat{k},\bar{k};\hat{p},\bar{p})}{\prod_{i=1}^{I}P_{2s}^{%
\prime }(\hat{k},\bar{k};\hat{p}_{i},\bar{p}_{i})}\right] , 
\]
where $I$ denotes the number of propagators, $p_{i}$ are linear combinations
of the external momenta and the numerator collects both the vertices and the
polynomials $P_{r}$ of (\ref{uvbeha}). Consider first the integral contained
in the square braket. Here $\hat{k}$ can be treated as an external momentum.
The regularity of the propagator ensures that differentiating the integrand
with respect to $\hat{k}$ (or $\hat{p}_{i}$, or $\bar{p}_{i}$) a sufficient
number of times the $\bar{k}$-integral becomes convergent. Thus, the
divergent part of the $\bar{k}$-integral is a polynomial $Q$ in $\hat{k}$
(and $\hat{p}_{i}$, $\bar{p}_{i}$). However, 
\[
\int \frac{\mathrm{d}^{\hat{d}-\varepsilon _{1}}\hat{k}}{(2\pi )^{\hat{d}}}Q(%
\hat{k};\hat{p},\bar{p})=0 
\]
in dimensional regularization, because it is the integral of a polynomial.
Thus the (sub)divergence of the $\bar{k}$-integral is killed by the $\hat{k}$%
-integral. An analogous conclusion holds exchanging the roles of $\hat{k}$
and $\bar{k}$. The arguments can be generalized to higher loops after
including the counterterms corresponding to the proper subdiagrams.

In a more general regularization setting the absence of $\hat{k}$- and $\bar{%
k}$-subdivergences is proved as follows. The overall divergences of the $%
\hat{k}$-$\overline{k}$-integrals are subtracted, for example, by the first
terms of the ``weighted Taylor expansion'' around vanishing external momenta 
\cite{renolor}. When the regularity conditions stated above are fulfilled,
those counterterms automatically cure also the $\hat{k}$-subintegrals and
the $\overline{k}$-subintegrals. Indeed, the subintegrals cannot behave
worse than the full integrals over $\hat{k}$ and $\overline{k}$, because (%
\ref{ba}) ensures that the propagators tend to zero with maximal velocity
also in the subintegrals, the loop-integration measures grow less rapidly
and the vertices grow not more rapidly than in the $\hat{k}$-$\overline{k}$%
-integrals.

The scalar and fermion propagators are clearly regular. On the other hand, a
propagator of the form 
\[
\frac{\Lambda _{L}^{n-1}}{|\hat{k}||\overline{k}|^{n}} 
\]
is not, and could generate ``spurious subdivergences'' when $\hat{k}$ tends
to infinity at $\overline{k}$ fixed, or viceversa. The problem appears in
certain large $N$ fermion models \cite{confnolor}, and becomes crucial
whenever gauge fields are present, as we now discuss.

The propagators (\ref{pros}) and (\ref{prosg}) are regular at non-vanishing
momenta, because the conditions (\ref{pos}) and (\ref{gfpos}) ensure that
the denominators are positive-definite in the Euclidean framework. To have
the best ultraviolet behaviors we must strengthen those conditions requiring
also 
\begin{equation}
\eta _{0}>0,\qquad \tau _{0}>0,\qquad \tilde{\eta}_{0}=\eta _{0}+\xi
_{0}>0,\qquad \zeta _{0}>0,  \label{must}
\end{equation}
which we assume from now on. However, attention must be paid to the
behaviors of propagators when $\hat{k}$ is sent to infinity at fixed $%
\overline{k}$, and when $\overline{k}$ is sent to infinity at fixed $\hat{k}$%
.

The conditions (\ref{must}) ensure that all gauge and ghost propagators are
regular in the Feynman gauge (\ref{fg})-(\ref{fg2}), except $\langle \bar{A}%
\bar{A}\rangle $, which has\ $\bar{\omega}\neq 0$, but $\hat{\omega}=0$, so
it is regular when $\overline{k}$ tends to infinity at $\hat{k}$ fixed, but
not when $\hat{k}$ tends to infinity at $\overline{k}$ fixed: in that region
of momentum space $\langle \bar{A}\bar{A}\rangle $ behaves like $\sim 1/\hat{%
k}^{2}$. To ensure that no spurious subdivergences are generated by the $%
\hat{k}$-subintegrals, we have to perform a more careful analysis, which is
done in appendix B and generalized in paper II. The result is that the
spurious subdivergences can be proved to be absent for 
\begin{equation}
\hat{d}=1,\qquad \text{\dj }<2+\frac{2}{n},\qquad d=\text{even},\qquad n=%
\text{odd,}  \label{subd}
\end{equation}
which we are going to assume from now on, unless explicitly stated. Observe
that the case (\ref{subd}) is a physically interesting one, since spacetime
is split into space and time. In four dimensions, (\ref{subd}) are
equivalent to just $\hat{d}=1$, $n=$odd (at $n>1$).

The absence of spurious subdivergences ensures the locality of counterterms.
Consider a diagram $G_{r}$ equipped with the subtractions that take care of
its diverging proper subdiagrams. Differentiating $G_{r}$ a sufficient
number of times with respect to any components $\hat{p}_{i}$, $\bar{p}_{i}$
of the external momenta $p_{i}$, we can arbitrarily reduce the overall
degree of divergence and eventually produce a convergent integral.
Therefore, overall divergences are polynomial in all components of the
external momenta.

\paragraph{Weighted power counting}

In the rest of this section we consider renormalizable and
super-renormalizable theories. We postpone the analysis of
strictly-renormalizable theories, namely the theories that contain only
weightless parameters, to section 8.

A generic vertex of (\ref{sur}) has the structure 
\begin{equation}
\lambda _{i}g^{n_{i}-2}\hat{\partial}^{k}\bar{\partial}^{m}\hat{A}^{p}\bar{A}%
^{q}\bar{C}^{r}C^{r}\varphi ^{s}\bar{\psi}^{t}\psi ^{t},  \label{vert}
\end{equation}
where $n_{i}=p+q+2r+s+2t$ and $p,q,r,k,m,s$ and $t$ are integers. Formula (%
\ref{vert}) and analogous expressions in this paper are meant
``symbolically'', which means that we pay attention to the field- and
derivative-contents of the vertices, but not where the derivatives act and
how Lorentz, gauge and other indices are contracted.

Every counterterm generated by (\ref{vert}) fits into the structure (\ref
{vert}). Indeed, consider a $L$-loop diagram with $E$ external legs, $I$
internal legs and $v_{i}$ vertices of type $i$. Counting the legs we have $%
\sum_{i}n_{i}v_{i}=E+2I=E+2(L+V-1)$, so the diagram is multiplied by a
product of couplings 
\begin{equation}
g^{\sum_{i}(n_{i}-2)v_{i}}\prod_{i}\lambda _{i}^{v_{i}}=\alpha
^{L}g^{E-2}\prod_{i}\lambda _{i}^{v_{i}}.  \label{ax}
\end{equation}
We see that a $g^{E-2}$ factorizes, as expected. Moreover, each loop order
carries an additional weight of at least $2[g]=4-$\dj .

\bigskip

We have already mentioned that when conditions (\ref{subd}) are fulfilled we
do not need to worry about the $\hat{k}$- and the $\overline{k}$%
-subintegrals, so we concentrate on the $\hat{k}$-$\overline{k}$-integrals.
The positivity conditions (\ref{pos}) and (\ref{gfpos}) ensure that the
denominators appearing in the Euclidean propagators (\ref{pros}) and (\ref
{prosg}) never vanish at non-exceptional momenta. Thus we have to study the
integrals only in the ultraviolet and infrared regions. We begin with the
ultraviolet behavior. We show that the $1/\alpha $ theories that satisfy (%
\ref{ono}) are renormalizable by weighted power counting, assuming that
spurious subdivergences are absent and that gauge (BRST) invariance is
preserved. In the next subsection we discuss the infrared behavior, while in
the next section we prove the preservation of BRST\ invariance.

In the analysis of renormalization based on weighted power counting, we can
treat the weightful parameters $\eta _{i}$, $\tau _{i}$, $\xi _{i}$ and $%
\zeta _{i}$, $i>0$, perturbatively, because the divergent parts of Feynman
diagrams depend polynomially on them. Then the propagators are (\ref{pros})
and (\ref{prosg}) with the replacements 
\[
\eta \rightarrow \eta _{0}\left( \frac{\bar{k}^{2}}{\Lambda _{L}^{2}}\right)
^{n-1},\qquad \tau \rightarrow \tau _{0}\left( \frac{\bar{k}^{2}}{\Lambda
_{L}^{2}}\right) ^{2(n-1)},\qquad \xi \rightarrow \xi _{0}\left( \frac{\bar{k%
}^{2}}{\Lambda _{L}^{2}}\right) ^{n-2},\qquad \zeta \rightarrow \zeta
_{0}\left( \frac{\bar{k}^{2}}{\Lambda _{L}^{2}}\right) ^{n-1}. 
\]
and every other term is treated as a vertex. Intermediate masses can be
added to the denominators, to avoid IR\ problems, and removed immediately
after calculating the divergent parts. Recall that (\ref{must}) are assumed.

The propagators have weights 
\begin{equation}
\lbrack \langle \hat{A}\hat{A}\rangle ]=[\langle C\bar{C}\rangle ]=[\langle
\varphi \varphi \rangle ]=-2,\qquad [\langle \psi \bar{\psi}\rangle
]=-1,\qquad [\langle \hat{A}\bar{A}\rangle ]=-3+\frac{1}{n},\qquad [\langle 
\bar{A}\bar{A}\rangle ]=-4+\frac{2}{n}.  \label{arg}
\end{equation}

Consider the vertex (\ref{vert}). Since the weight of (\ref{vert}) must be
equal to \dj , we have the inequality 
\begin{equation}
n_{i}+k+\frac{m+q}{n}-q+t\leq 4.  \label{fr}
\end{equation}
The degree of divergence $\omega (G)$ of a diagram $G$ with $L$ loops, $\hat{%
I}$, $\tilde{I}$ and $\bar{I}$ internal legs of type $\hat{A}\hat{A}$, $\hat{%
A}\bar{A}$ and $\bar{A}\bar{A}$, respectively, $I_{\text{gh}}$, $I_{\varphi
} $ and $I_{\psi }$ internal ghost, scalar and fermion legs, $\hat{E}$, $%
\bar{E}$, $E_{\text{gh}}$, $E_{\varphi }$ and $E_{\psi }$ external $\hat{A}$%
-, $\bar{A}$-, ghost, scalar and fermion legs, and $v_{i}$ vertices (\ref
{vert}), where $i$ stands for $(k,m,p,q,r,s,t)$, is 
\[
\omega (G)=L\text{\dj }-2(\hat{I}+I_{\text{gh}}+I_{\varphi })-I_{\psi
}-\left( 3-\frac{1}{n}\right) \tilde{I}-\left( 4-\frac{2}{n}\right) \bar{I}%
+\sum_{i}v_{i}\left( k+\frac{m}{n}\right) . 
\]
Using $L=I-V+1$, (\ref{fr}) and the leg-countings 
\begin{eqnarray*}
\hat{E}+2\hat{I}+\tilde{I} &=&\sum_{i}pv_{i},\qquad \bar{E}+2\bar{I}+\tilde{I%
}=\sum_{i}qv_{i},\qquad E_{\text{gh}}+2I_{\text{gh}}=\sum_{i}2rv_{i}, \\
E_{\varphi }+2I_{\varphi } &=&\sum_{i}sv_{i},\qquad E_{\psi }+2I_{\psi
}=\sum_{i}2tv_{i},
\end{eqnarray*}
we get 
\begin{equation}
\omega (G)\leq \text{\dj }-(\hat{E}+E_{\text{gh}}+E_{\varphi })\left( \frac{%
\text{\dj }}{2}-1\right) -\bar{E}\left( \frac{\text{\dj }}{2}-2+\frac{1}{n}%
\right) -E_{\psi }\left( \frac{\text{\dj }-1}{2}\right) -(4-\text{\dj }%
)\left( L+\frac{E}{2}-1\right) ,  \label{boh}
\end{equation}
where $E$ is the total number of external legs. The divergent part of $G$ is
a weighted polynomial of degree $\omega (G)$ in the external momenta.
Recalling (\ref{ax}), the counterterms have the form (\ref{vert}), precisely 
\[
\left( \prod_{i}\lambda _{i}^{n_{i}}\right) \alpha ^{L}g^{E-2}\hat{\partial}%
^{K}\bar{\partial}^{M}\hat{A}^{\hat{E}}\bar{A}^{\bar{E}}(\bar{C}C)^{E_{\text{%
gh}}/2}\varphi ^{E_{\varphi }}(\bar{\psi}\psi )^{E_{\psi }/2},\qquad \text{%
with }K+\frac{M}{n}=\omega (G), 
\]
and therefore are subtracted renormalizing an appropriate coupling $\lambda
_{i}$. This proves that the theory is renormalizable by weighted power
counting assuming that BRST\ invariance is preserved and that the spurious
subdivergences are absent.

\paragraph{Absence of infrared divergences}

Now we study the infrared behavior of correlations functions. Correlation
functions have to be understood as distributions and calculated off-shell at
non-exceptional external momenta. Exceptional configurations of momenta can
be reached by analytical continuation. However, the continuation exists if
the theory does not contain super-renormalizable vertices and massless
fields, otherwise infrared divergences occur in Feynman diagrams at high
orders, even off-shell \cite{apple}. We work out the conditions under which
such problems do not occur in our models.

All theories with a non-trivial super-renormalizable subsector look alike at
low energies. Up to terms of higher dimensions, they are described by the
lagrangian (for generic $\hat{d}$) 
\begin{equation}
\mathcal{L}_{\text{IR}}=\frac{1}{4}\left[ (F_{\hat{\mu}\hat{\nu}%
}^{a})^{2}+2\eta _{n-1}(F_{\hat{\mu}\bar{\nu}}^{a})^{2}+\tau _{2n-2}(F_{\bar{%
\mu}\bar{\nu}}^{a})^{2}\right] ,  \label{ir}
\end{equation}
where $\eta _{n-1}$ and $\tau _{2n-2}$ are constants. Moreover the
gauge-fixing becomes $\mathcal{G}_{\text{IR}}^{a}=\hat{\partial}\cdot \hat{A}%
^{a}+\zeta _{n-1}\bar{\partial}\cdot \bar{A}^{a}$. We assume, compatibly
with (\ref{pos}) and (\ref{gfpos}), 
\[
\eta _{n-1}>0,\qquad \tau _{2n-2}>0,\qquad \zeta _{n-1}>0. 
\]

In general, the theory $\mathcal{L}_{\text{IR}}$ is Lorentz violating, but
has an ordinary power counting, so we can use the results known from
ordinary quantum field theory, which tell us that the Feynman diagrams of $%
\mathcal{L}_{\text{IR}}$ have no IR divergences (at non-exceptional external
momenta) if and only if 
\begin{equation}
d\geq 4.  \label{irabse}
\end{equation}
Observe that if $d>4$ the theory $\mathcal{L}_{\text{IR}}$ is
non-renormalizable, but this fact does not concern our present discussion,
since our models are not just $\mathcal{L}_{\text{IR}}$, but contain terms
of higher dimensionalities that cure the UV behavior.

In strictly renormalizable theories different conditions apply (see section
6).

Massless fields are responsible for another type of IR divergences, those
that occur in cross sections of non-confining gauge theories (infrared
catastrophe). Such divergences are usually treated with the Bloch-Nordsieck
resummation method \cite{bloch}. This issue is beyond the scope of the
present paper, yet we expect that the argument of Bloch and Nordsieck can be
adapted also to Lorentz violating theories.

\bigskip

In conclusion, recalling (\ref{ono}), (\ref{subd}) and (\ref{irabse}), the
conditions to have consistent renormalizable $1/\alpha $ gauge theories with
a non-trivial super-renormalizable subsector are 
\begin{equation}
d=\text{even}\geq 4,\qquad \hat{d}=1,\qquad \text{\dj }<2+\frac{2}{n},\qquad
n=\text{odd}.  \label{d=42}
\end{equation}
In four dimensions they reduce to just $\hat{d}=1$, $n=$odd (for $n>1$).

\section{Renormalizability to all orders}

\setcounter{equation}{0}

So far we have concentrated on the renormalizability of our theories by
weighted power counting. It remains to prove that the subtraction of
divergences is compatible with gauge invariance. We use the
Batalin-Vilkovisky formalism \cite{batalin}. For simplicity, we concentrate
on pure gauge theories and use the minimal subtraction scheme and the
dimensional-regularization technique. In particular, the functional
integration measure is automatically BRST\ invariant.

Classical proofs of the renormalizability of (Lorentz invariant) Yang-Mills
theories can be found in most textbooks \cite{weinbergetal}. Complete
classifications of the BRST\ cohomology of local operators \cite{joglekar}
and local functionals of arbitrary ghost number \cite{barnich} are
available. The generalization of such classification theorems to Lorentz
violating theories appears to be conceptually simple, but technically
involved, and is beyond the scope of this paper.

\paragraph{Batalin-Vilkovisky formalism}

The fields are collectively denoted by $\Phi ^{i}=(A_{\mu }^{a},\overline{C}%
^{a},C^{a},B^{a})$. Add BRST sources $K_{i}=(K_{a}^{\mu },K_{\overline{C}%
}^{a},K_{C}^{a},K_{B}^{a})$ for every field $\Phi ^{i}$ and extend the
action (\ref{basis}) as 
\begin{equation}
\Sigma (\Phi ,K)=\mathcal{S}(\Phi )-\int \mathrm{d}^{d}x\left[ \left(
sA_{\mu }^{a}\right) K_{a}^{\mu }+\left( s\overline{C}^{a}\right) K_{%
\overline{C}}^{a}+\left( sC^{a}\right) K_{C}^{a}+\left( sB^{a}\right)
K_{B}^{a}\right] ,  \label{acca}
\end{equation}
From (\ref{acca}) we can read the weights of the BRST\ sources: 
\begin{equation}
\lbrack K_{a}^{\hat{\mu}}]=[K_{\overline{C}}^{a}]=[K_{C}^{a}]=\frac{\text{%
\dj }}{2},\qquad [K_{a}^{\bar{\mu}}]=\frac{\text{\dj }}{2}+1-\frac{1}{n}%
,\qquad [K_{B}^{a}]=\frac{\text{\dj }}{2}-1.  \label{iss}
\end{equation}
Define the antiparenthesis 
\begin{equation}
(X,Y)=\int \mathrm{d}^{d}x\left\{ \frac{\delta _{r}X}{\delta \Phi ^{i}(x)}%
\frac{\delta _{l}Y}{\delta K_{i}(x)}-\frac{\delta _{r}X}{\delta K_{i}(x)}%
\frac{\delta _{l}Y}{\delta \Phi ^{i}(x)}\right\} .  \label{antipar}
\end{equation}
BRST\ invariance is generalized to the identity 
\begin{equation}
(\Sigma ,\Sigma )=0,  \label{nil}
\end{equation}
which is a straightforward consequence of (\ref{acca}), the gauge invariance
of $\mathcal{S}_{0}$ and the nilpotency of $s$. Define also the generalized
BRST operator 
\begin{equation}
\sigma X\equiv (\Sigma ,X),  \label{sigma}
\end{equation}
which is nilpotent ($\sigma ^{2}=0$), because of the identity (\ref{nil}).
Observe that $\sigma $, as well as $s$, raises the weight by one unit.

The generating functionals $Z$, $W$ and $\Gamma $ are defined, in the
Euclidean framework, as 
\begin{eqnarray}
Z[J,K] &=&\int \mathcal{D}\Phi \exp \left( -\Sigma (\Phi ,K)+\int \Phi
^{i}J_{i}\right) =\text{e}^{W[J,K]},  \label{zj} \\
\Gamma [\Phi _{\Gamma },K] &=&-W[J,K]+\int \Phi _{\Gamma }^{i}J_{i},\qquad 
\text{where\qquad }\Phi _{\Gamma }^{i}=\frac{\delta _{r}W[J,K]}{\delta J_{i}}%
.  \nonumber
\end{eqnarray}
Below we often suppress the subscript $\Gamma $ in $\Phi _{\Gamma }$.
Performing a change of variables 
\begin{equation}
\Phi ^{\prime }=\Phi +\theta s\Phi ,  \label{chv}
\end{equation}
in the functional integral (\ref{zj}), $\theta $ being a constant
anticommuting parameter, and using the identity (\ref{nil}), we find 
\begin{equation}
(\Gamma ,\Gamma )=0.  \label{milpo}
\end{equation}

A canonical transformation of fields and sources is defined as a
transformation that preserves the antiparenthesis. It is generated by a
functional $\mathcal{F}(\Phi ,K^{\prime })$ and reads 
\[
\Phi ^{i\ \prime }=\frac{\delta \mathcal{F}}{\delta K_{i}^{\prime }},\qquad
K_{i}=\frac{\delta \mathcal{F}}{\delta \Phi ^{i}}. 
\]
The generating functional of the identity transformation is 
\[
\mathcal{I}(\Phi ,K^{\prime })=\int \mathrm{d}^{d}x\sum_{i}\Phi
^{i}K_{i}^{\prime }. 
\]

As usual, renormalizability is proved proceeding inductively. The inductive
assumption is that up to the $n$-th loop included the divergences can be
removed redefining the physical parameters $\alpha _{i}$ and performing a
canonical transformation of the fields and BRST sources. Call $\Sigma _{n}$
and $\Gamma ^{(n)}$ the action and generating functional renormalized up to
the $n$-th loop included. The inductive assumption ensures that $\Sigma _{n}$
and $\Gamma ^{(n)}$ satisfy (\ref{nil}) and (\ref{milpo}), respectively.

Locality and (\ref{milpo}) imply that the ($n+1$)-loop divergences $\Gamma
_{n+1\ \text{div}}^{\ (n)}$ of $\Gamma ^{(n)}$ are local and $\sigma $%
-closed, namely $\sigma \Gamma _{n+1\ \text{div}}^{\ (n)}=0$. We have to
work out the most general solution to this condition. First, observe that $%
\Gamma _{n+1\ \text{div}}^{\ (n)}$ cannot contain $B^{a}$, $K_{B}^{a}$ and $%
K_{\overline{C}}^{a}$, because the action (\ref{acca}) provides no vertices
with $B^{a}$, $K_{B}^{a}$ or $K_{\overline{C}}^{a}$ on the external legs. In
particular, the absence of vertices with $B$-legs is due to the linearity of
the gauge-fixing $\mathcal{G}^{a}$ (\ref{gf}) in the gauge field $A$.
Second, observe that the vertices that contain an antighost $\bar{C}$
contain also a $\hat{\partial}$ or a $\zeta \left( \bar{\upsilon}\right) 
\bar{\partial}$ acting on $\bar{C}$. The vertex containing $\hat{\partial}%
\overline{C}$ has an identical vertex-partner where $\hat{\partial}\overline{%
C}$ is replaced by $\hat{K}_{A}$, while the vertex containing $\zeta \left( 
\bar{\upsilon}\right) \bar{\partial}\overline{C}$ has an identical
vertex-partner where $\zeta \left( \bar{\upsilon}\right) \bar{\partial}%
\overline{C}$ is replaced by $\bar{K}_{A}$. Therefore $\Gamma _{n+1\ \text{%
div}}^{\ (n)}$ can depend on $\overline{C}$, $\hat{K}_{A}$ and $\bar{K}_{A}$
only through the combinations 
\[
K_{a}^{\hat{\mu}}+\partial ^{\hat{\mu}}\overline{C}^{a},\qquad K_{a}^{\bar{%
\mu}}+\zeta \left( \bar{\upsilon}\right) \partial ^{\bar{\mu}}\overline{C}%
^{a}. 
\]

Using the facts just proved, invariance under global gauge transformations
and the weighted power counting, we find that in the $1/\alpha $ theories $%
\Gamma _{n+1\ \text{div}}^{\ (n)}$ has the form 
\begin{eqnarray}
\Gamma _{n+1\ \text{div}}^{\ (n)} &=&\int \mathrm{d}^{d}x\left[ \widetilde{%
\mathcal{G}}_{n}(A)+\left( K_{a}^{\hat{\mu}}+\partial ^{\hat{\mu}}\overline{C%
}^{a}\right) \left( a_{n}^{\prime }\partial _{\hat{\mu}}C^{a}+h_{n}^{abc}A_{%
\hat{\mu}}^{b}C^{c}\right) \right.  \nonumber \\
&&\left. +\left( K_{a}^{\bar{\mu}}+\zeta \left( \bar{\upsilon}\right)
\partial ^{\bar{\mu}}\overline{C}^{a}\right) \left( b_{n}^{\prime }\partial
_{\bar{\mu}}C^{a}+k_{n}^{abc}A_{\bar{\mu}}^{b}C^{c}\right)
+e_{n}gf^{abc}K_{C}^{a}C^{b}C^{c}\right] ,  \label{ss0}
\end{eqnarray}
where $\widetilde{\mathcal{G}}_{n}$ depends only on $A_{\mu }^{a}$ and has
weight \dj , while $a_{n}^{\prime }$, $b_{n}^{\prime }$, $e_{n}$, $%
h_{n}^{abc}$ and $k_{n}^{abc}$ are weightless constants. Considering the
terms proportional to $(\partial ^{\hat{\mu}}\overline{C}^{a})(\partial _{%
\hat{\mu}}C^{b})C^{c}$ and $(\partial ^{\bar{\mu}}\overline{C}^{a})(\partial
_{\bar{\mu}}C^{b})C^{c}$ contained in $\sigma \Gamma _{\text{div}}^{\
(n+1)}=0$, we see that $h_{n}^{abc}$ and $k_{n}^{abc}$ must be proportional
to $f^{abc}$. Then (\ref{ss0}) can be reorganized in the more convenient
form 
\begin{eqnarray}
\Gamma _{n+1\ \text{div}}^{\ (n)} &=&\int \mathrm{d}^{d}x\left[ \widetilde{%
\mathcal{G}}_{n}(A)+\left( K_{a}^{\hat{\mu}}+\partial ^{\hat{\mu}}\overline{C%
}^{a}\right) \left( a_{n}\partial _{\hat{\mu}}C^{a}+c_{n}D_{\hat{\mu}%
}C^{a}\right) \right.  \nonumber \\
&&\left. +\left( K_{a}^{\bar{\mu}}+\zeta \left( \bar{\upsilon}\right)
\partial ^{\bar{\mu}}\overline{C}^{a}\right) \left( b_{n}\partial _{\bar{\mu}%
}C^{a}+d_{n}D_{\bar{\mu}}C^{a}\right)
+e_{n}gf^{abc}K_{C}^{a}C^{b}C^{c}\right] ,  \label{simpe}
\end{eqnarray}
with new constants $a_{n}$, $b_{n}$, $c_{n}$ and $d_{n}$. Working out the
condition $\sigma \Gamma _{\text{div}}^{\ (n+1)}=0$ in detail we find $%
c_{n}=d_{n}=2e_{n}$ and 
\[
\widetilde{\mathcal{G}}_{n}=\mathcal{G}_{n}-a_{n}\frac{\delta \mathcal{S}_{0}%
}{\delta A_{\hat{\mu}}^{a}}A_{\hat{\mu}}^{a}-b_{n}\frac{\delta \mathcal{S}%
_{0}}{\delta A_{\bar{\mu}}^{a}}A_{\bar{\mu}}^{a}, 
\]
where $\mathcal{G}_{n}$ is gauge invariant and $\mathcal{S}_{0}$ is given by
formula (\ref{s0}). We have used the property that the gauge-field equations 
$\delta \mathcal{S}_{0}/\delta A_{\hat{\mu}}^{a}$ and $\delta \mathcal{S}%
_{0}/\delta A_{\bar{\mu}}^{a}$ transform covariantly. The result can be
collected into the compact form 
\begin{equation}
\Gamma _{n+1\ \text{div}}^{\ (n)}=\int \mathrm{d}^{d}x\left( \mathcal{G}%
_{n}+\sigma \mathcal{R}_{n}\right) ,  \label{counter}
\end{equation}
with 
\begin{equation}
\mathcal{R}_{n}(\Phi ,K)=\int \mathrm{d}^{d}x\left(
-a_{n}I_{1}-b_{n}I_{2}+c_{n}I_{3}\right) ,  \label{ter}
\end{equation}
where 
\begin{equation}
I_{1}(\Phi ,K)=(K_{a}^{\hat{\mu}}+\partial ^{\hat{\mu}}\overline{C}^{a})A_{%
\hat{\mu}}^{a},\qquad I_{2}(\Phi ,K)=\left( K_{a}^{\bar{\mu}}+\partial ^{%
\bar{\mu}}\zeta \left( \bar{\upsilon}\right) \overline{C}^{a}\right) A_{\bar{%
\mu}}^{a},\qquad I_{3}(\Phi ,K)=K_{C}^{a}C^{a},  \label{abasis}
\end{equation}

Now,$\mathcal{\ G}_{n}$ is local, gauge-invariant, constructed with $A$ and
its derivatives, and has weight \dj . Since, by assumption, $\mathcal{S}_{0}$
contains the full set of such terms, $\mathcal{G}_{n}$ can be reabsorbed
renormalizing the physical couplings $\alpha _{i}$ contained in $\mathcal{S}%
_{0}$. We denote these renormalization constants by $Z_{\alpha _{i}}$.

The $\sigma $-exact counterterms are reabsorbed with a canonical
transformation generated by 
\begin{equation}
\mathcal{F}_{n}(\Phi ,K^{\prime })=\mathcal{I}(\Phi ,K^{\prime })-\mathcal{R}%
_{n}(\Phi ,K^{\prime }).  \label{iui}
\end{equation}
More explicitly, $\overline{C}^{a}$, $B^{a}$ and $K_{B}^{a}$ are
non-renormalized and the only non-trivial redefinitions are 
\begin{eqnarray}
\hat{A}^{a} &\rightarrow &\hat{Z}_{n\hspace{0.01in}A}^{1/2}\hat{A}%
^{a},\qquad \bar{A}^{a}\rightarrow \bar{Z}_{n\hspace{0.01in}A}^{1/2}\bar{A}%
^{a},\qquad C^{a}\rightarrow Z_{n\hspace{0.01in}C}^{1/2}C^{a},\qquad
K_{C}^{a}\rightarrow Z_{n\hspace{0.01in}C}^{-1/2}K_{C}^{a},  \nonumber \\
K_{a}^{\hat{\mu}} &\rightarrow &\hat{Z}_{n\hspace{0.01in}A}^{-1/2}(K_{a}^{%
\hat{\mu}}+\partial ^{\hat{\mu}}\overline{C}^{a})-\partial ^{\hat{\mu}}%
\overline{C}^{a},\qquad K_{a}^{\bar{\mu}}\rightarrow \bar{Z}_{n\hspace{0.01in%
}A}^{-1/2}(K_{a}^{\bar{\mu}}+\zeta \left( \bar{\upsilon}\right) \partial ^{%
\bar{\mu}}\overline{C}^{a})-\zeta \left( \bar{\upsilon}\right) \partial ^{%
\bar{\mu}}\overline{C}^{a},  \nonumber \\
\qquad K_{\overline{C}}^{a} &\rightarrow &K_{\overline{C}}^{a}+\hat{\partial}%
\cdot \hat{A}^{a}(\hat{Z}_{n\hspace{0.01in}A}^{1/2}-1)+\zeta \left( \bar{%
\upsilon}\right) \bar{\partial}\cdot \bar{A}^{a}(\bar{Z}_{n\hspace{0.01in}%
A}^{1/2}-1),  \label{refeda}
\end{eqnarray}
where 
\begin{equation}
\hat{Z}_{n\hspace{0.01in}A}^{1/2}=1+a_{n},\qquad \bar{Z}_{n\hspace{0.01in}%
A}^{1/2}=1+b_{n},\qquad Z_{n\hspace{0.01in}C}^{1/2}=1-c_{n}.  \label{warda}
\end{equation}

Call $f(Z_{n})$ the map (\ref{refeda}), where $Z_{n}=(\hat{Z}_{nA},\bar{Z}%
_{nA},Z_{nC})$. It is straightforward to check that it satisfies the group
property 
\begin{equation}
f(Z_{p})\circ f(Z_{q})=f(Z_{p}Z_{q}),\qquad Z_{p}Z_{q}\equiv (\hat{Z}_{pA}%
\hat{Z}_{qA},\bar{Z}_{pA}\bar{Z}_{qA},Z_{pC}Z_{qC}).  \label{group}
\end{equation}
The ($n+1$)-loop divergences (\ref{counter}) are reabsorbed by a map $%
h(Z_{n\alpha },Z_{n})$ obtained composing the renormalizations of the
physical couplings $\alpha _{i}$ with $f(Z_{n})$. Clearly, also $h$
satisfies the group property (\ref{group}). Moreover, the basis (\ref{abasis}%
) is $h$- and $f$-invariant.

Being a composition of a canonical transformation and redefinitions of the
physical couplings, $h(Z_{n\alpha },Z_{n})$ preserves both (\ref{nil}) and (%
\ref{milpo}). This proves the inductive hypothesis to the order $n+1$, and
therefore promotes (\ref{nil}) and (\ref{milpo}) to all orders.

The complete renormalization of divergences is performed by the map 
\[
h_{\infty }=\prod_{n=1}^{\infty }h(Z_{n\alpha },Z_{n})=h(Z_{\alpha
},Z),\qquad Z_{\alpha }=\prod_{n=1}^{\infty }Z_{n\alpha },\qquad
Z=\prod_{n=1}^{\infty }Z_{n}. 
\]
Applying $h_{\infty }$\ to the action $\Sigma $, it is easy to prove that
the renormalization can be equivalently performed in a more standard
multiplicative fashion, namely 
\begin{eqnarray}
\hat{A}^{a} &\rightarrow &\hat{Z}_{\hspace{0.01in}A}^{1/2}\hat{A}^{a},\qquad 
\bar{A}^{a}\rightarrow \bar{Z}_{\hspace{0.01in}A}^{1/2}\bar{A}^{a},\qquad
C^{a}\rightarrow Z_{\hspace{0.01in}C}^{1/2}C^{a},\qquad \overline{C}%
^{a}\rightarrow \hat{Z}_{\hspace{0.01in}A}^{-1/2}\overline{C}^{a},  \nonumber
\\
B^{a} &\rightarrow &\hat{Z}_{\hspace{0.01in}A}^{-1/2}B^{a},\qquad \lambda
\rightarrow \lambda \hat{Z}_{\hspace{0.01in}A},\qquad \zeta \rightarrow \hat{%
Z}_{\hspace{0.01in}A}^{1/2}\bar{Z}_{A}^{-1/2}\zeta ,\qquad \alpha
_{i}\rightarrow \alpha _{i}Z_{\alpha _{i}}.  \label{cota}
\end{eqnarray}
The renormalization constants of the BRST\ sources are the reciprocals of
the renormalization constants of the fields: $\Phi ^{i}\rightarrow \Phi
^{i}Z_{i}^{1/2}\Leftrightarrow K_{i}\rightarrow K_{i}Z_{i}^{-1/2}$.

The relatively simple structure of the counterterms is due to the simple
form of the gauge fixing (\ref{gf}), which is linear in the gauge potential.
Had we chosen a non-linear gauge fixing, for example the most general local
function $\mathcal{G}^{a}$ of weight 2 constructed with the gauge potential
and its derivatives, there would be vertices with $B$-external legs, and
therefore also counterterms with $B$'s on the external legs. Then $\mathcal{R%
}_{n}$ would be much less constrained, to allow for the most general
renormalization of the gauge-fixing parameters contained in $\mathcal{G}^{a}$%
.

At the practical level, computations considerably simplify using the
background field method \cite{background}, which can be applied
straightforwardly to our theories.

\section{Renormalizable theories}

\setcounter{equation}{0}

In this section we investigate the four dimensional renormalizable theories
and study the low-energy recovery of Lorentz invariance.

We start with the $1/\alpha $ theory with $\hat{d}=1$, $n=2$, \dj $=5/2$.
Its (Euclidean) lagrangian reads
\begin{eqnarray}
\mathcal{L}_{1/\alpha } &=&\frac{1}{4}\left[ \frac{2\eta _{0}}{\Lambda
_{L}^{2}}(D_{\bar{\rho}}^{ab}F_{\hat{\mu}\bar{\nu}}^{b})^{2}+\frac{\tau _{0}%
}{\Lambda _{L}^{4}}(\bar{D}^{2}F_{\bar{\mu}\bar{\nu}}^{a})^{2}+\frac{\xi _{0}%
}{\Lambda _{L}^{2}}(D_{\hat{\rho}}F_{\bar{\mu}\bar{\nu}})(D_{\hat{\rho}}F_{%
\bar{\mu}\bar{\nu}})+2\eta _{1}(F_{\hat{\mu}\bar{\nu}}^{a})^{2}\right.  
\nonumber \\
&&\left. +\frac{\tau _{1}}{\Lambda _{L}^{2}}(D_{\bar{\rho}}^{ab}F_{\bar{\mu}%
\bar{\nu}}^{b})^{2}+\tau _{2}(F_{\bar{\mu}\bar{\nu}}^{a})^{2}\right] +\frac{g%
}{\Lambda _{L}^{2}}f_{abc}\left( \lambda F_{\hat{\mu}\bar{\nu}}^{a}F_{\hat{%
\mu}\bar{\rho}}^{b}+\lambda ^{\prime }F_{\bar{\mu}\bar{\nu}}^{a}F_{\bar{\mu}%
\bar{\rho}}^{b}\right) F_{\bar{\nu}\bar{\rho}}^{c}  \nonumber \\
&&+\frac{g}{\Lambda _{L}^{4}}\sum_{j}\lambda _{j}\bar{D}^{2}\bar{F}%
^{3}{}_{j}+\frac{\alpha }{\Lambda _{L}^{4}}\sum_{k}\lambda _{k}^{\prime }%
\bar{F}^{4}{}_{k},  \label{de}
\end{eqnarray}
where $j$ labels the independent gauge invariant terms constructed with two
covariant derivatives $\bar{D}$ acting on three field strengths $\bar{F}$,
and $k$ labels the terms constructed with four $\bar{F}$'s. The last two
terms are symbolic, while the rest contains the precise list of allowed
terms.

In every super-renormalizable $1/\alpha $ theory we have $\beta _{\Lambda
_{L}}=\beta _{\tau _{0}}=\beta _{\eta _{0}}=\beta _{\xi _{0}}=\beta _{\alpha
}=0$. Indeed, we know that $\Lambda _{L}$ is RG invariant, by construction.
So are $\eta _{0}$, $\tau _{0}$ and $\xi _{0}$, because they are weightless.
The $\alpha $-beta function vanishes, because each vertex carries at least a
factor of $g$.

The parameters of the model (\ref{de}) are actually finite. Indeed, we have
the weights $[g]=3/4$, $[\eta _{1}]=[\tau _{1}]=[\lambda ^{\prime }]=1$, $%
[\tau _{2}]=2$, $[\lambda ]=[\lambda _{j}]=[\lambda _{k}^{\prime }]=0$ and
by (\ref{ax}) every diagram is multiplied by $\alpha ^{L}g^{E-2}$.
Therefore, no counterterm can fit into the structure (\ref{de}). However,
the model (\ref{de}) has an even $n$, so it may have spurious
subdivergences. Going through the analysis of Appendix B it is possible to
show that such subdivergences appear only at three loops.

The low energy limit of (\ref{de}) can be studied taking $\Lambda _{L}$ to
infinity. We get 
\begin{equation}
\mathcal{L}_{\text{IR}}=\frac{1}{4}\left[ 2\eta _{1}(F_{\hat{\mu}\bar{\nu}%
}^{a})^{2}+\tau _{2}(F_{\bar{\mu}\bar{\nu}}^{a})^{2}\right] .  \label{lew}
\end{equation}
Lorentz invariance is recovered, because the redefinition 
\begin{eqnarray}
\hat{x}^{\prime } &=&\hat{x},\qquad \hat{A}^{\prime }=(\eta _{1}\tau
_{2})^{1/4}\hat{A},\qquad \bar{x}^{\prime }=\eta _{1}^{1/2}\tau _{2}^{-1/2}%
\bar{x},\qquad \bar{A}^{\prime }=\eta _{1}^{-1/4}\tau _{2}^{3/4}\bar{A}, 
\nonumber \\
\alpha ^{\prime } &=&(\eta _{1}\tau _{2})^{-1/2}\alpha ,\qquad \mu ^{\prime
}=\mu ,\qquad \Lambda _{L}^{\prime }=\eta _{1}^{-1}\tau _{2}\Lambda _{L},
\label{relo}
\end{eqnarray}
converts the low-energy action into the manifestly Lorentz invariant form 
\begin{equation}
\mathcal{S}_{\text{IR}}=\int \mathrm{d}^{4}x\ \mathcal{L}_{\text{IR}}=\int 
\mathrm{d}^{4}x^{\prime }\ \frac{1}{4}(F_{\mu \nu }^{a\ \prime })^{2}.
\label{low}
\end{equation}

In (\ref{relo}) we have included the $\mu $- and $\Lambda _{L}$%
-transformations so the redefinition can be applied also to the high-energy
theory, if the remaining couplings are rescaled appropriately. The
divergences of the low-energy theory include those due to the $\Lambda
_{L}\rightarrow \infty $ limit. Moreover, to have Lorentz invariance at low
energies, the (low-energy) subtraction scheme has to be properly adjusted.

Observe that (\ref{relo}) is a combination of a gauge-field normalization, 
\begin{equation}
A^{\prime }=\eta _{1}\tau _{2}^{-1/2}A,\qquad \alpha ^{\prime }=\eta
_{1}^{-2}\tau _{2}\alpha ,  \label{norma}
\end{equation}
a usual rescaling, 
\begin{equation}
x^{\prime }=\eta _{1}\tau _{2}^{-1}x,\qquad A^{\prime }=\eta _{1}^{-1}\tau
_{2}A,\qquad \alpha ^{\prime }=\alpha ,\qquad \mu ^{\prime }=\eta
_{1}^{-1}\tau _{2}\mu ,\qquad \Lambda _{L}^{\prime }=\eta _{1}^{-1}\tau
_{2}\Lambda _{L},  \label{relo2}
\end{equation}
and a weighted rescaling \cite{renolor}, 
\begin{eqnarray}
\hat{x}^{\prime } &=&\eta _{1}^{-1}\tau _{2}\hat{x},\qquad \hat{A}^{\prime
}=\eta _{1}^{1/4}\tau _{2}^{-1/4}\hat{A},\qquad \bar{x}^{\prime }=\eta
_{1}^{-1/2}\tau _{2}^{1/2}\bar{x},\qquad \bar{A}^{\prime }=\eta
_{1}^{-1/4}\tau _{2}^{1/4}\bar{A},  \nonumber \\
\alpha ^{\prime } &=&\eta _{1}^{3/2}\tau _{2}^{-3/2}\alpha ,\qquad \mu
^{\prime }=\eta _{1}\tau _{2}^{-1}\mu ,\qquad \Lambda _{L}^{\prime }=\Lambda
_{L}.  \label{relo3}
\end{eqnarray}
Moreover, note that (\ref{relo}) leaves $\mu $ unchanged (although it
changes $\Lambda _{L}$), therefore it generates no anomalous contributions
at high energies. On the other hand, anomalous effects are generated at low
energies. Because of (\ref{relo2}) and (\ref{relo3}), such effects are the
difference between the trace anomaly and the weighted trace anomaly, and
correspond to a $\Lambda _{L}$-running with no $\mu $-running. They amount
to a low-energy scheme change and are taken into account in the scheme
adjustment mentioned above, necessary to restore Lorentz invariance at low
energies. The gauge-fixing sector of the theory can remain Lorentz violating
with no observable consequence.

\bigskip

In general, to recover Lorentz invariance at low energies we are free to
perform just one redefinition, which amounts, ultimately, to a particular $%
\bar{x}$-rescaling. In the case (\ref{lew}) the Lorentz recovery is possible
thanks to the simple structure of the theory. Consider a more general
situation, for example the theory (\ref{lew}) coupled to fermions. Its
lagrangian is $\mathcal{L}_{gf}=\mathcal{L}_{1/\alpha }+\mathcal{L}_{f}$,
where 
\begin{equation}
\mathcal{L}_{f}=\bar{\psi}\left( \hat{D}\!\!\!\!\slash+\frac{\eta _{0f}}{%
\Lambda _{L}}{\bar{D}\!\!\!\!\slash}^{\,2}+\eta _{1f}{\bar{D}\!\!\!\!\slash}%
+m_{f}+\frac{\tau _{f}g}{\Lambda _{L}}iT^{a}{\bar{F}}_{\bar{\mu}\bar{\nu}%
}^{a}\sigma ^{\bar{\mu}\bar{\nu}}\right) \psi .  \label{de2}
\end{equation}
The low-energy lagrangian reads now 
\begin{equation}
\mathcal{L}_{\text{IR }gf}=\frac{1}{4}\left[ 2\eta _{1}(F_{\hat{\mu}\bar{\nu}%
}^{a})^{2}+\tau _{2}(F_{\bar{\mu}\bar{\nu}}^{a})^{2}\right] +\bar{\psi}(\hat{%
D}\!\!\!\!\slash+\eta _{1f}\bar{D}{\!\!\!\!\slash+m_{f})}\psi .  \label{low2}
\end{equation}
The transformations (\ref{norma}), (\ref{relo2}) and (\ref{relo3}) extend to
fermions as 
\[
\psi ^{\prime }=\psi ,\qquad \psi ^{\prime }=\eta _{1}^{-3/2}\tau
_{2}^{3/2}\psi ,\qquad \psi ^{\prime }=\eta _{1}^{3/4}\tau _{2}^{-3/4}\psi , 
\]
respectively, so 
\[
\int \mathrm{d}^{4}x\ \mathcal{L}_{\text{IR }gf}=\int \mathrm{d}%
^{4}x^{\prime }\ \left[ \frac{1}{4}(F_{\mu \nu }^{a\ \prime })^{2}+\bar{\psi}%
^{\prime }\left( {\hat{D}\!\!\!\!\slash}^{\,\prime }+\eta _{1f}^{\prime }{%
\bar{D}\!\!\!\!\slash}^{\,\prime }+m_{f}\right) \psi ^{\prime }\right] , 
\]
where 
\[
\eta _{1f}^{\prime }=\eta _{1f}\left( \frac{\eta _{1}}{\tau _{2}}\right)
^{1/2}. 
\]
Therefore, Lorentz invariance cannot be recovered at low energies unless the
low-energy couplings are located on the ``Lorentz invariant surface'' 
\begin{equation}
\tau _{2}=\eta _{1f}^{2}\eta _{1}.  \label{le}
\end{equation}
If that does not happen the effects of the violation become observable also
at low energies.

The Lorentz invariant surface (\ref{le}) is also RG invariant, which means
that there exists a low-energy subtraction scheme such that the beta
function of $\delta _{L}\equiv \tau _{2}-\eta _{1f}^{2}\eta _{1}$ is
proportional to $\delta _{L}$. RG invariance guarantees that it is not
necessary to specify at which low-energy scale the relation (\ref{le}) must
hold: if it holds at some low-energy scale it holds at all low-energy
scales. Perturbative calculations suggest \cite{nielsen} that in
CPT-invariant theories the Lorentz invariant\ surface might also be RG\
stable, i.e. the couplings that parametrize the displacement from the
surface, such as $\delta _{L}$, are IR free\footnote{%
Other results \cite{colladay} give evidence that CPT-violating couplings
exhibit the opposite behavior.}. When the Lorentz invariant surface is not
RG\ stable, Lorentz invariance can be recovered at low energies only by
means of an appropriate fine-tuning. With great accuracy experiments show 
\cite{koste2} that the Standard Model is located on the Lorentz invariant
surface.

If we look at the issue of Lorentz-invariance recovery from a low-energy
viewpoint, it is more natural to consider (\ref{de}) plus (\ref{de2}),
equipped with (\ref{le}), as a (partial) regularization of the
Lorentz-invariant Yang-Mills theory 
\[
\int \mathrm{d}^{4}x\left[ \frac{1}{4}(F_{\mu \nu }^{a})^{2}+\bar{\psi}%
\left( {D\!\!\!\!\slash}+m_{f}\right) \psi \right] .
\]
In this description, the existence of a Lorentz-preserving subtraction
scheme is obvious and the fine-tuning (\ref{le}) appears more natural. Then,
however, low-energy Lorentz invariance is assumed from the start. Moreover,
we insist that our theories should not be viewed as regularization devices,
but as true, fundamental theories, to be experimentally tested.

The model (\ref{de}) is not free of spurious subdivergences, because $n$ is
even. The first completely consistent model is thus the $1/\alpha $ theory
with $\hat{d}=1$, $n=3$, \dj $=2$, which is studied in detail in paper II.
Its simplest renormalizable lagrangian is the sum of the quadratic part (\ref
{l0}) plus $\bar{F}^{3}$. Other consistent four dimensional solutions to (%
\ref{d=42}) exist for every odd $n\geq 3$. The low-energy considerations of
this section are very general and apply to the odd-$n$ models with obvious
modifications.

\section{Strictly renormalizable theories}

\setcounter{equation}{0}

For the sake of completeness, we investigate strictly renormalizable and
weighted scale invariant theories. We recall that there exists a class of
subtraction schemes in which no power-like divergences are generated. In
those schemes super-renormalizable parameters are not turned on by
renormalization. Moreover, such class of schemes is automatically chosen by
the dimensional-regularization technique, which we assume in this paper.

In strictly-renormalizable theories the quadratic part $\mathcal{L}_{Q}$ of
the lagrangian must have 
\[
\eta (\bar{\Upsilon})=\eta _{0}\bar{\Upsilon}^{n-1},\qquad \tau (\bar{%
\Upsilon})=\tau _{0}\bar{\Upsilon}^{2(n-1)},\qquad \xi (\bar{\Upsilon})=\xi
_{0}\bar{\Upsilon}^{n-2}. 
\]
For convenience we can choose a strictly-renormalizable gauge fixing, with $%
\zeta (\bar{\upsilon})=\zeta _{0}\bar{\upsilon}^{n-1}$. The inequalities (%
\ref{must}) must hold.

The IR behavior of Feynman diagrams is still dominated by the weighted power
counting, so the analysis of potential IR\ divergences differs from the one
of section 3. Now $\eta (0)=\tau (0)=0$, so the gauge-field propagator
contains additional denominators $\sim 1/\bar{k}^{2(n-1)}$ in the $%
\left\langle \bar{A}\bar{A}\right\rangle $-sector. The loop integrals over $%
k $ and the loop sub-integrals over $\bar{k}$ are IR divergent unless 
\begin{equation}
\text{\dj }>4-\frac{2}{n},\qquad \bar{d}>2(n-1),  \label{stric}
\end{equation}
respectively. The former condition follows from (\ref{arg}). The latter
condition and $n\geq 2$ imply $\bar{d}\geq 3$.

We can now distinguish two cases: if \dj $=4$ the gauge coupling is
strictly-renormalizable, while if \dj $<4$ the theory can be
strictly-renormalizable only if it is Abelian, but not $1/\alpha $. However,
the models with \dj $=4$ do not satisfy (\ref{subd}), so we cannot ensure
that they are free of subdivergences, even at $\hat{d}=1$. Moreover, such
models exist only in dimensions greater than $6$. Indeed, it is easy to see
that $n\geq 2$, \dj $=4$ and (\ref{subd}) imply 
\[
\bar{d}=3n,\qquad d=1+3n\geq 7, 
\]
and (\ref{stric}) are automatically satisfied. The simplest example of this
kind is the seven-dimensional theory with $n=2$, $\bar{d}=6$. Its lagrangian
reads, in symbolic form 
\begin{equation}
\mathcal{L}=\mathcal{L}_{Q}+\frac{\lambda }{\Lambda _{L}^{7/2}}\tilde{F}^{2}%
\bar{F}+\frac{\lambda ^{\prime }}{\Lambda _{L}^{11/2}}\bar{D}^{2}\bar{F}^{3}+%
\frac{\lambda ^{\prime \prime }}{\Lambda _{L}^{7}}\bar{F}^{4}.  \label{oh}
\end{equation}

Other examples of strictly renormalizable theories are the fixed points of
the RG flow, which are exactly weighted scale invariant. Following \cite
{confnolor}, we can attempt to construct four dimensional weighted scale
invariant theories in the large $N$ expansion, where $N$ is the number of
fermion copies. However, we can easily prove that this attempt fails in the
presence of gauge fields. Consider the model with lagrangian 
\begin{equation}
\mathcal{L}=\sum_{i=1}^{N}\bar{\psi}_{i}\left( \hat{D}\!\!\!\!\slash+\frac{%
\bar{D}{\!\!\!\!\slash\,}^{n}}{\Lambda _{L}^{n-1}}\right) \psi _{i}.
\label{n1}
\end{equation}
For simplicity we assume that the gauge field is Abelian, but the argument
generalizes straightforwardly to non-Abelian gauge fields. The gauge fields
do not have a kinetic term, which is provided by a one-loop diagram. Since
the gauge-field propagator is dynamically generated, the regularity
conditions (\ref{subd}) might have to be replaced by more sophisticated
restrictions. We assume (\ref{stric}) to ensure that Feynman diagrams are
free of IR divergences. We study under which conditions the theory (\ref{n1}%
) is renormalizable in the form (\ref{n1}). Observe that if $\bar{d}>1$
there always exist counterterms of the form 
\begin{equation}
\frac{1}{\Lambda _{L}^{n-1}}\sum_{i=1}^{N}\bar{\psi}_{i}F_{\bar{\mu}\bar{\nu}%
}\sigma ^{\bar{\mu}\bar{\nu}}\bar{D}{\!\!\!\!\slash\,}^{n-2}\psi _{i},
\label{t1}
\end{equation}
that make the theory (\ref{n1}) not renormalizable. Then we must require $%
\bar{d}=1$, so that the counterterms (\ref{t1}) are trivial. However, this
condition is incompatible with (\ref{stric}).

We conclude that Lorentz violating gauge theories with strictly
renormalizable gauge couplings are problematic at $n>1$.

\section{Proca theories}

\setcounter{equation}{0}

We conclude considering Proca versions of our theories and study the UV
behaviors of their propagators. Instead of being gauge-fixed, now the
lagrangian (\ref{l0}) is equipped with a mass term 
\[
\mathcal{L}_{m}=\frac{m^{2}}{2}\left( \hat{A}^{2}+\bar{A}\tilde{\zeta}\left( 
\bar{\upsilon}\right) \bar{A}\right) 
\]
for the gauge fields, where $\widetilde{\zeta }$ is a polynomial of degree $%
n-1$. At the free-field level, acting with a derivative on the field
equations we get the identity 
\[
\hat{\partial}\cdot \hat{A}+\tilde{\zeta}\left( \bar{\upsilon}\right) \bar{%
\partial}\cdot \bar{A}=0, 
\]
which kills one degree of freedom.

Expressing the propagator of $\mathcal{L}_{Q}+\mathcal{L}_{m}$ in the form (%
\ref{pros}) we find 
\begin{eqnarray*}
u &=&\frac{1}{D(1,\eta )+m^{2}},\qquad v=\frac{1}{D(\tilde{\eta},\tau )+m^{2}%
\tilde{\zeta}},\qquad r=\frac{\eta }{m^{2}\left( \eta D(1,\tilde{\zeta}%
)+m^{2}\tilde{\zeta}\right) }, \\
s &=&\frac{ur}{\eta }\left( \eta D(1,\eta )+m^{2}\tilde{\zeta}\right)
,\qquad t=\frac{vr}{\eta }\left( \eta D(\tilde{\eta},\tau )+m^{2}\tilde{\tau}%
\right) .
\end{eqnarray*}
While $u$ and $v$ have regular behaviors, we have 
\[
r,s,t\sim \frac{1}{m^{2}\hat{k}^{2}},\quad \text{for }\hat{k}^{2}\rightarrow
\infty ,\qquad r,s,t\sim \frac{1}{m^{2}\tilde{\zeta}\bar{k}^{2}},\quad \text{%
for }\bar{k}^{2}\rightarrow \infty . 
\]
We see that the large-momentum behaviors of $u$ and $v$ agree with weighted
power counting, but those of $r$, $s$ and $t$ dot not. We conclude that the
Lorentz violating Proca theories are not renormalizable.

\section{Conclusions}

\setcounter{equation}{0}

In this paper we have constructed Lorentz violating gauge theories that can
be renormalized by weighted power counting. The theories contain higher
space derivatives, but are arranged so that no counterterms with higher time
derivatives are generated by renormalization. The absence of spurious
subdivergences privileges the models where spacetime is split into space and
time. We have focused on the simplest class of models, leaving the general
classification of renormalizable theories to a separate paper.

If Lorentz invariance is violated at high energies there remains to explain
why it should be recovered at low energies, since generically
renormalization makes the couplings run independently and there is no
apparent reason why the parameters of the low-energy theory should belong to
the Lorentz invariant surface. It is of course possible to restore Lorentz
invariance at low energies by means of a fine tuning, which is easier to
justify when the Lorentz invariant surface is RG stable.

\vskip 20truept \noindent {\Large \textbf{Acknowledgments}}

\vskip 10truept

I am grateful to P. Menotti for drawing my attention to ref.s \cite{nielsen}
and for discussions. I thank the referee for stimulating remarks.

\vskip 20truept \noindent {\Large \textbf{Appendix A: Classification of the
quadratic terms}}

\vskip 10truept

\renewcommand{\theequation}{A.\arabic{equation}} \setcounter{equation}{0}

In this appendix we derive the form (\ref{l0}) of the quadratic lagrangian $%
\mathcal{L}_{Q}$. It contains all terms of weights $\leq $\dj , constructed
with two field strengths and possibly covariant derivatives. Clearly there
exists a single such term with two $\hat{F}$'s, that is $\hat{F}_{\mu \nu
}^{2}$. Consider now the terms constructed with two $\bar{F}$'s and possibly
derivatives $\bar{D}$. We start from 
\begin{equation}
F_{\bar{\mu}\bar{\nu}}D_{\bar{\lambda}}\cdots D_{\bar{\tau}}F_{\bar{\rho}%
\bar{\sigma}}  \label{op}
\end{equation}
and study all possible contractions. Up to addition of vertices, we can
freely commute the covariant derivatives, since $[D_{\bar{\alpha}},D_{\bar{%
\beta}}]=gF_{\bar{\alpha}\bar{\beta}}$. From now on every formula of this
appendix is meant up to vertices and total derivatives. Clearly, contracting
four derivatives of (\ref{op}) with the indices $\bar{\mu},\bar{\nu},\bar{%
\rho},\bar{\sigma}$ we obtain a vertex. Contracting two derivatives with
field-strength indices we obtain 
\[
\mathcal{I}_{p}\equiv F_{\bar{\mu}\bar{\nu}}(\bar{D}^{2})^{p}D_{\bar{\rho}%
}D_{\bar{\mu}}F_{\bar{\rho}\bar{\nu}}.
\]
Using the Bianchi identity on $D_{\bar{\mu}}F_{\bar{\rho}\bar{\sigma}}$ we
easily get 
\[
\mathcal{I}_{p}=\frac{1}{2}F_{\bar{\mu}\bar{\nu}}(\bar{D}^{2})^{p+1}F_{\bar{%
\mu}\bar{\nu}}.
\]
This is the unique independent contraction of (\ref{op}). By weighted power
counting, $p$ can be at most $2n-3$. The terms of this type are those
corresponding to the function $\tau $ of (\ref{l0}).

Next, consider terms with two $\bar{F}$'s and derivatives $\bar{D}$ and $%
\hat{D}$. By $O(1,\hat{d}-1)$ invariance these terms can contain two $\hat{D}
$'s, which must be contracted among themselves, or no $\hat{D}$, which is
the case already considered. Arguing as before, we get new terms of the form 
\[
\mathcal{I}_{p}^{\prime }=\frac{1}{2}F_{\bar{\mu}\bar{\nu}}\hat{D}^{2}(\bar{D%
}^{2})^{p}F_{\bar{\mu}\bar{\nu}},\qquad p\leq n-2, 
\]
which correspond to the function $\xi $ in (\ref{l0}).

The terms with one $\bar{F}$ and one $\tilde{F}$ are 
\[
\mathcal{I}_{p}^{\prime \prime }=F_{\hat{\mu}\bar{\nu}}D_{\hat{\mu}}(\bar{D}%
^{2})^{p}D_{\bar{\sigma}}F_{\bar{\nu}\bar{\sigma}}. 
\]
Using the Bianchi identities we get $\mathcal{I}_{p}^{\prime \prime }=-%
\mathcal{I}_{p}^{\prime }$, therefore nothing new. The terms with two $%
\tilde{F}$'s are 
\[
\mathcal{I}_{p}^{\prime \prime \prime }=F_{\hat{\mu}\bar{\nu}}D_{\bar{\lambda%
}_{1}}\cdots D_{\bar{\lambda}_{2p+2}}F_{\hat{\mu}\bar{\sigma}}. 
\]
No $\hat{D}$ derivatives can have place here, by weighted power counting.
Contracting two derivatives with the field-strength indices $\bar{\nu}$ and $%
\bar{\sigma}$ we get the unique new contraction 
\begin{equation}
\mathcal{I}_{p}^{\prime \prime \prime }-\mathcal{I}_{p}^{\prime \prime }=F_{%
\hat{\mu}\bar{\nu}}(\bar{D}^{2})^{p+1}F_{\hat{\mu}\bar{\nu}},\qquad p\leq
n-2,  \label{stop}
\end{equation}
which gives the terms corresponding to $\eta $ in (\ref{l0}). Finally, it is
easy to see that no term with one $\hat{F}$ and one $\tilde{F}$, or one $%
\hat{F}$ and one $\bar{F}$, are allowed by weighted power counting.

\vskip 20truept \noindent {\Large \textbf{Appendix B: Absence of spurious
subdivergences}}

\vskip 10truept

\renewcommand{\theequation}{B.\arabic{equation}} \setcounter{equation}{0}

In this appendix we study the spurious subdivergences. We first need to
classify the lagrangian terms that contain derivatives $\hat{\partial}$. We
know that no term can contain more than two $\hat{\partial}$'s. Moreover,
vertices with two $\hat{\partial}$'s cannot contain fermions and ghosts,
because their weights would exceed \dj . For the same reason, they cannot
contain more than one scalar and do not depend on $\hat{A}$. Moreover,
vertices with one $\hat{\partial}$ cannot contain fermions. Summarizing, $%
\hat{\partial}$-dependent vertices have necessarily the forms 
\begin{equation}
X_{1}\equiv \hat{\partial}f_{1}(\hat{A},\bar{A},\varphi ,\bar{C},C,\bar{%
\partial}),\quad \quad X_{2}\equiv f_{2}(\bar{A},\varphi ,\bar{\partial})(%
\hat{\partial}\bar{A})(\hat{\partial}\bar{A}),\quad \quad X_{2}^{\prime
}\equiv f_{2}^{\prime }(\bar{A},\varphi ,\bar{\partial})(\hat{\partial}^{2}%
\bar{A}),  \label{vk1}
\end{equation}
where the $\bar{\partial}$-derivatives are allowed to act anywhere, as well
as the $\hat{\partial}$-derivative in $X_{1}$. The functions $f_{1}$, $f_{2}$
and $f_{2}^{\prime }$ are constrained by locality, weighted power counting
and BRST invariance, and their structure depends on $n$. However, their form
is not relevant for the proof that follows.

The quadratic terms that do not fall in the classes (\ref{vk1}) are 
\begin{equation}
(\hat{\partial}\hat{A})^{2},\qquad \bar{C}\hat{\partial}^{2}C,\qquad \varphi 
\hat{\partial}^{2}\varphi ,\qquad \bar{\psi}\hat{\partial}\psi .
\label{quad}
\end{equation}
Every other lagrangian term is $\hat{\partial}$-independent.

Our purpose is to derive sufficient conditions to ensure that all integrals
are free of subdivergences, once counterterms for proper divergent
subdiagrams are included. In particular, spurious subdivergences must be
absent, because they do not correspond to any subdiagram, so there exist no
counterterms that can subtract them. We work in the Feynman gauge (\ref{fg}%
)-(\ref{fg2}) and assume that the spacetime dimension is even, together with 
$\hat{d}=1$, $n=$odd. We use the dimensional regularization and proceed
inductively in the loop order. The proof is considerably involved, and we
have to split it in three steps.

\paragraph{First step: structure of integrals}

Consider a generic $N$-loop integral

\begin{equation}
\int \frac{\mathrm{d}\hat{k}_{1}}{(2\pi )^{\hat{d}}}\int \frac{\mathrm{d}^{%
\bar{d}}\bar{k}_{1}}{(2\pi )^{\bar{d}}}\cdots \int \frac{\mathrm{d}\hat{k}%
_{N}}{(2\pi )^{\hat{d}}}\int \frac{\mathrm{d}^{\bar{d}}\bar{k}_{N}}{(2\pi )^{%
\bar{d}}},  \label{a1}
\end{equation}
with loop momenta $(k_{1},\ldots ,k_{N})$. We have to prove that all
subintegrals, in all parametrizations $(k_{1}^{\prime },\ldots
,k_{N}^{\prime })$ of the momenta, are free of subdivergences. By the
inductive assumption, all subintegrals 
\begin{equation}
\prod_{j=1}^{M}\int \frac{\mathrm{d}\hat{k}_{j}^{\prime }}{(2\pi )^{\hat{d}}}%
\int \frac{\mathrm{d}^{\bar{d}}\bar{k}_{j}^{\prime }}{(2\pi )^{\bar{d}}},
\label{a2}
\end{equation}
where $M<N$, are subtracted, if divergent, by appropriate counterterms. We
need to consider integrals where some hatted integrations are missing and
the corresponding barred integrations are present, and/or viceversa. Start
from subintegrals $\mathcal{\bar{I}}$ containing (\ref{a2}) and one integral 
\begin{equation}
\int \frac{\mathrm{d}^{\bar{d}}\bar{k}_{a}^{\prime }}{(2\pi )^{\bar{d}}}
\label{a3}
\end{equation}
for some $a$, but no integration over $\hat{k}_{a}^{\prime }$. The form of
propagators (\ref{fg2}) and (\ref{prosg}) ensures that differentiating $%
\mathcal{\bar{I}}$ a sufficient number of times with respect to $\hat{k}%
_{a}^{\prime }$ the subintegral $\mathcal{\bar{I}}$ becomes overall
convergent. Thus, its overall (spurious) $\mathcal{\bar{I}}$-subdivergence
is polynomial in $\hat{k}_{a}^{\prime }$. However, in the complete integral (%
\ref{a1}) $\mathcal{\bar{I}}$ must eventually be integrated over $\hat{k}%
_{a}^{\prime }$. This operation kills the spurious subdivergence, because in
dimensional regularization the integral of a polynomial vanishes.

Next, consider subintegrals $\mathcal{\bar{I}}$ containing (\ref{a2}) and a
product 
\[
\prod_{a}\int \frac{\mathrm{d}^{\bar{d}}\bar{k}_{a}^{\prime }}{(2\pi )^{\bar{%
d}}}, 
\]
but no integration over the corresponding $\hat{k}_{a}^{\prime }$s. Then
there exists a combination of derivatives 
\[
\prod_{a}\frac{\partial ^{n_{a}}}{\partial \hat{k}_{a}^{\prime \ n_{a}}}, 
\]
for suitable $n_{a}$s, that cures not only the overall divergence of $%
\mathcal{\bar{I}}$, but also its subdivergences (e.g. those of type (\ref{a3}%
)). Thus, the spurious subdivergences are a combination of contributions,
each of which is local in at least one $\hat{k}_{a}^{\prime }$: again, such
spurious subdivergences are killed by the integrals over the $\hat{k}%
_{a}^{\prime }$s.

We see that we do not need to worry about subintegrals $\mathcal{\bar{I}}$
containing an excess of barred integrations. On the other hand, we do need
to worry about subintegrals $\mathcal{\hat{I}}$ containing an excess of
hatted integrations or excesses of both types. We start from the
subintegrals containing only hatted integrations.

\paragraph{Second step: $\hat{k}$-subintegrals}

Consider the $\hat{k}$-subintegral of a diagram $G$ with $L$ loops, $v_{1}$
vertices of type $X_{1}$, $v_{2}$ vertices of type $X_{2}$ and $%
X_{2}^{\prime }$, $\Delta v$ vertices of other types, $I_{B}$ internal
bosonic legs (including ghosts) and $I_{F}$ internal fermionic legs. We have
seen in section 3 that every bosonic propagator behaves at least like $1/%
\hat{k}^{2}$, for $\hat{k}$ large, while the fermionic propagator behaves
like $1/\hat{k}$. The $\hat{k}$-subintegral behaves like 
\begin{equation}
\int \mathrm{d}^{L\hat{d}}\hat{k}\ \frac{\hat{k}^{v_{1}+2v_{2}}}{(\hat{k}%
^{2})^{I_{B}}\hat{k}^{I_{F}}},\qquad \hat{\omega}(G)=L\hat{d}%
+v_{1}+2v_{2}-2I_{B}-I_{F},  \label{bint}
\end{equation}
$\hat{\omega}(G)$ denoting its degree of divergence. Using $%
L=1+I_{B}+I_{F}-v_{1}-v_{2}-\Delta v$, we can write 
\[
\hat{\omega}(G)=L(\hat{d}-2)+2+\Delta \hat{\omega}(G),\qquad \Delta \hat{%
\omega}(G)=I_{F}-v_{1}-2\Delta v.
\]
We know that no vertices with four or more fermionic legs are allowed in $%
1/\alpha $ theories. Because of this fact, the fermionic internal lines must
end at different vertices of the set $\Delta v$. Therefore, we have $%
I_{F}\leq \Delta v$ and $\Delta \hat{\omega}(G)\leq 0$. Then the integral (%
\ref{bint}) is certainly convergent for $\hat{d}=1$ and $L>2$. Instead, for $%
\hat{d}>1$ there exist divergent diagrams with arbitrarily many loops.

From now on we assume $\hat{d}=1$. We must consider the one- and two-loop
diagrams more explicitly. Observe that no logarithmic divergence exists at
one loop in odd (i.e. $\hat{d}$) dimensions. More precisely, 
\[
\int \mathrm{d}^{\hat{d}}\hat{k}\ \frac{\hat{k}_{\hat{\mu}_{1}}\cdots \hat{k}%
_{\hat{\mu}_{2m-1}}}{(\hat{k}^{2})^{m}} 
\]
is UV convergent by symmetric integration. Here we have kept $\hat{d}$
generic to emphasize that it is continued to complex values. Moreover,
power-like divergences are absent using the dimensional-regularization
technique.

We remain only with two-loop diagrams. Setting $\hat{d}=1$ and $L=2$ we find 
$\hat{\omega}(G)=I_{F}-v_{1}-2\Delta v\leq -v_{1}-\Delta v\leq 0$. The
potential divergence is logarithmic and can only occur for $%
I_{F}=v_{1}=\Delta v=0$. Let us leave the quadratic terms (\ref{quad}) aside
for a moment. Divergent diagrams can contain only vertices of types $X_{2}$
and $X_{2}^{\prime }$. Moreover, the $\hat{\partial}$'s of $X_{2}$ and $%
X_{2}^{\prime }$ must act on internal legs and the external momenta can be
set to zero. The internal legs can only be of type $\bar{A}\bar{A}$, plus
possibly one internal $\varphi $-leg. At $L=2$ we have $v_{2}+1$ internal
legs, so the diagram has the form of Fig. 1, (a) or (b). However, diagram
(a) is the product of two one-loop diagrams, so it does not diverge.
Consider now the vertex F of diagram (b). It can be an $X_{2}$ or an $%
X_{2}^{\prime }$. The two $\hat{k}$'s shown in the figure are those
belonging to F. If F is an $X_{2}$, the vertex and the two propagators
attached to it make in total $\hat{k}\cdot \hat{k}/(\hat{k}^{2})^{2}=1/\hat{k%
}^{2}$. The same conclusion holds if F is an $X_{2}^{\prime }$. Therefore,
the divergent part of diagram (b) coincides (apart from external factors)
with the one of the modified diagram where F is suppressed. Similarly, we
can suppress the vertices A, B, C, D, E and G, and reduce to the diagram of
Fig. 1, (b$^{\prime }$), which has the form 
\begin{equation}
\int \mathrm{d}^{\hat{d}}\hat{k}\ \mathrm{d}^{\hat{d}}\hat{p}\ \frac{P_{4}(%
\hat{k},\hat{p})}{\hat{k}^{2}\hat{p}^{2}(\hat{k}+\hat{p})^{2}},  \label{into}
\end{equation}
where $P_{4}(\hat{k},\hat{p})$ is a scalar degree-4 polynomial in $\hat{k}$
and $\hat{p}$. Clearly, $P_{4}(\hat{k},\hat{p})$ can also be written as a
degree-2 polynomial $P_{2}$ in $\hat{k}^{2}$, $\hat{p}^{2}$ and $(\hat{k}+%
\hat{p})^{2}$. We see that $P_{2}$ is a linear combination of terms, each of
which simplifies at least one denominator of (\ref{into}), leaving a sum of
integrals of the form 
\[
\int \mathrm{d}^{\hat{d}}\hat{k}\ \mathrm{d}^{\hat{d}}\hat{p}\ \frac{a\hat{k}%
^{2}+b\hat{p}^{2}+c\hat{k}\cdot \hat{p}}{\hat{k}^{2}\hat{p}^{2}} 
\]
(possibly after a $\hat{k}$- or $\hat{p}$-translation), where $a$, $b$ and $%
c $ are constants. The first two contributions are zero in dimensional
regularization, while the third contribution factorizes into the product of
two one-loop integrals, which cannot have logarithmic divergences for the
reasons explained before.

The two-leg vertices (\ref{quad}) leave $\hat{\omega}(G)$ unchanged. On the
other hand, we have seen that potentially divergent diagrams have only $\bar{%
A}$- or $\varphi $-internal legs, so only the scalar term of (\ref{quad})
can be used. However, it simplifies a scalar propagator, so we end up again
with (\ref{into}). 
\begin{figure}[tbp]
\centerline{\includegraphics[width=4in,height=1.5in]{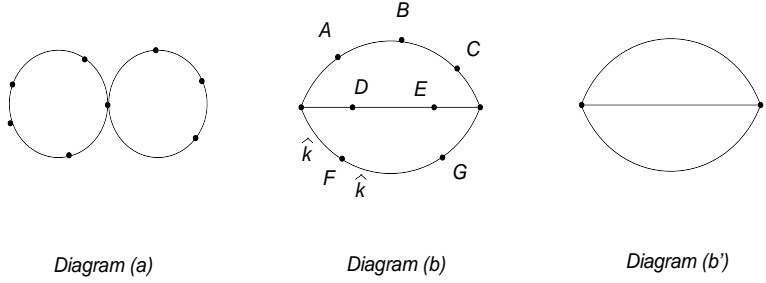}}
\caption{Analysis of the spurious subdivergences}
\end{figure}

\paragraph{Third step: mixed subintegrals}

Now we consider subintegrals of the form 
\begin{equation}
\prod_{i=1}^{L}\int \frac{\mathrm{d}\hat{k}_{i}^{\prime }}{(2\pi )^{\hat{d}}}%
\left[ \prod_{j=L+1}^{L+M}\int \frac{\mathrm{d}\hat{k}_{j}^{\prime }}{(2\pi
)^{\hat{d}}}\int \frac{\mathrm{d}^{\bar{d}}\bar{k}_{j}^{\prime }}{(2\pi )^{%
\bar{d}}}\right] ,  \label{a4}
\end{equation}
which are ``incomplete'' in $L$ barred directions. The complete subintegrals
in square brakets can be regarded as products of (nonlocal, but one-particle
irreducible) ``subvertices''. Let $\tilde{v}_{r}$ be the number of
subvertices of type $r$, with $\tilde{n}_{\hat{A}r}$, $\tilde{n}_{\bar{A}r}$%
, $\tilde{n}_{Cr}$, $\tilde{n}_{fr}$, $\tilde{n}_{sr}$ external legs of
types $\hat{A}$, $\bar{A}$, ghost, fermion and scalar, respectively. Since
subvertices are at least one-loop, each leg has a factor $g$ attached to it
(see (\ref{ax})). Thus, the weight $\tilde{\delta}_{r}$ of the subvertex of
type $r$ satisfies the bound 
\begin{equation}
\tilde{\delta}_{r}\leq \text{\dj }-\tilde{n}_{\hat{A}r}-\frac{\tilde{n}_{%
\bar{A}r}}{n}-\tilde{n}_{Cr}-\frac{3}{2}\tilde{n}_{fr}-\tilde{n}_{sr}.
\label{deltatilde}
\end{equation}
Now we repeat the argument leading to (\ref{bint}) for diagrams that may
contain such subvertices. We have 
\begin{equation}
\hat{\omega}(G)=L+v_{1}+2v_{2}-2I_{B}-I_{F}+\sum_{r}\tilde{v}_{r}\tilde{%
\delta}_{r}.  \label{b1}
\end{equation}
Moreover, the topological identity $L-I+V=1$ gives 
\begin{equation}
L=1+I_{B}+I_{F}-v_{1}-v_{2}-\Delta v-\sum_{r}\tilde{v}_{r}.  \label{b2}
\end{equation}
We can write $\Delta v=\Delta v_{B}+\Delta v_{F}$, to distinguish the $%
\Delta v_{B}$ vertices containing no fermionic legs from the $\Delta v_{F}$
vertices containing two fermionic legs. Counting the fermionic legs of our
subdiagram we have 
\begin{equation}
2I_{F}+E_{F}=2\Delta v_{F}+\sum_{r}\tilde{v}_{r}\tilde{n}_{fr},  \label{b3}
\end{equation}
where $E_{F}$ denotes the number of external fermionic legs. Combining (\ref
{b1}), (\ref{b2}) and (\ref{b3}) we get 
\begin{equation}
\hat{\omega}(G)=2-L-v_{1}-2\Delta v_{B}-\Delta v_{F}-\frac{E_{F}}{2}+\sum_{r}%
\tilde{v}_{r}\left( \tilde{\delta}_{r}-2+\frac{1}{2}\tilde{n}_{fr}\right) .
\label{degre}
\end{equation}
We know that in the realm of the usual power counting, odd-dimensional
integrals do not have logarithmic divergences. In the realm of the weighted
power counting, such a property generalizes as follows: if $\hat{d}=1$, $d=$%
even and $n=$odd, then odd-dimensional (weighted) integrals do not have
logarithmic divergences. The proof is simple and left to the reader. Thus,
the case $L=1$ is excluded. Sufficient conditions to have $\hat{\omega}%
(G)\leq 0$ are then 
\begin{equation}
\tilde{\delta}_{r}-2+\frac{1}{2}\tilde{n}_{fr}<0\qquad \text{for every }r%
\text{.}  \label{soll}
\end{equation}
Indeed, if (\ref{soll}) hold (\ref{degre}) gives $\hat{\omega}(G)<0$ unless
subvertices are absent, which is the case considered previously.

Finally, the most general mixed subintegrals have the form 
\[
\prod_{i=1}^{L}\int \frac{\mathrm{d}\hat{k}_{i}^{\prime }}{(2\pi )^{\hat{d}}}%
\left[ \prod_{j=L+1}^{L+M}\int \frac{\mathrm{d}\hat{k}_{j}^{\prime }}{(2\pi
)^{\hat{d}}}\int \frac{\mathrm{d}^{\bar{d}}\bar{k}_{j}^{\prime }}{(2\pi )^{%
\bar{d}}}\prod_{m=L+M+1}^{L+M+P}\int \frac{\mathrm{d}^{\bar{d}}\bar{k}%
_{m}^{\prime }}{(2\pi )^{\bar{d}}}\right] . 
\]
They can be treated as above, considering the integrals between square
brakets as subvertices. Now formula (\ref{degre}) has an extra $-P$ on the
right-hand side, since $P$ hatted intergations are missing. The situation,
therefore, can only improve. The only caveat is that now $L$ can also be one
(if $P$ is odd). Even in that case, however, $2-L-P\leq 0$, since $P\geq 1$.

\paragraph{Restrictions}

Using (\ref{deltatilde}), sufficient conditions for (\ref{soll}) are 
\[
\text{\dj }-\tilde{n}_{\hat{A}}-\frac{\tilde{n}_{\bar{A}r}}{n}-\tilde{n}%
_{Cr}-\tilde{n}_{fr}-\tilde{n}_{sr}<2. 
\]
The worst case is $\tilde{n}_{\bar{A}r}=2$, $\tilde{n}_{\hat{A}}=\tilde{n}%
_{Cr}=\tilde{n}_{fr}=\tilde{n}_{sr}=0$, which gives 
\[
\text{\dj }<2+\frac{2}{n}. 
\]
This condition is always satisfied in four dimensions (for $n>1$).
Summarizing, we have been able to prove the absence of spurious
subdivergences under the sufficient conditions (\ref{subd}).

\bigskip

Some final remarks are in order. The conclusions of this appendix do not
apply to the case $n=1$, because then $\tau $, $\eta $ and $\zeta $ are
constant and $\xi $ vanishes, so the propagators (\ref{pros}) are regular.
Thus for $n=1$ all types of Lorentz breakings are allowed, which is
well-known. Since (\ref{subd}) are sufficient, but not necessary,
conditions, we cannot exclude all models that violate them. In specific
cases other types of cancellations can take place, because of symmetries or
peculiar types of expansions or resummations (e.g. large $N$). Generically
speaking, even some theories with $\hat{d}>1$, $n>1$ might work, although we
are unable to give explicit examples of that kind right now.

\end{document}